\documentclass[aps,pra,onecolumn,showpacs]{revtex4}

\usepackage{physics,graphicx,amssymb,epstopdf}
\begin{document}

\title{Ion trap architectures and new directions}

\author{James D. Siverns$^1$ and Qudsia Quraishi$^{1,2}$}

\affiliation{$^1$Joint Quantum Institute, University of Maryland, College Park, MD, 20740  \\ $^{2}$Army Research Laboratory, 2800 Powder Mill Rd., Adelphi, MD, 20783}

\email{jsiverns@umd.edu}
\email{qudsia.quraishi@gmail.com}

\date{compiled \today}

\begin{abstract}
Trapped ion technology has seen advances in performance, robustness, and versatility over the last decade. With increasing numbers of trapped ion groups world-wide, a myriad of trap architectures are currently in use. Applications of trapped ions include: quantum simulation, computing and networking, time standards and fundamental studies in quantum dynamics. Design of such traps is driven by these various research aims, but some universally desirable properties have lead to the development of ion trap foundries. The excellent control achievable with trapped ions and the ability to do photonic-readout, has allowed progress on quantum networking using entanglement between remotely situated ion-based nodes. Here we present a selection of trap architectures currently in use by the community and present their most salient characteristics, identifying features particularly suited for quantum networking. We also discuss our own in-house research efforts aimed at long-distance trapped ion-networking.

%\keywords{03.67.-a \and 03.67.Hk \and 03.67.Lx}
\end{abstract}

\maketitle

\section{Introduction}
Since their development by Wolfgang Paul \cite{Paul90} and Hans Dehmelt \cite{Dehmelt67} in the 1950s and 1960s, ion traps have been an integral part of a wide range of experiments including frequency standards \cite{Bollinger91,Fisk97,Rosenband08,Huntemann16,Keller16,Chou1630}, mass spectrometry \cite{Schwartz2002}, cavity QED \cite{Keller03,Kreuter04,Barros09,Hiroki13} and fundamental physics \cite{Odom06} as well as quantum simulation \cite{Porras04,Porras06,Islam13,Schindler2013,Zhang17,Neyenhuis16}, quantum information processing \cite{CiracZoller95,Milburn00,MolmerSorensen00,Duan04,Wineland98,Debnath2016} and quantum networking \cite{Monroe14}. This wide range of applications is, in part, due to the ion trap's ability to provide good isolation of the trapped atom from the environment while allowing the user to manipulate the internal state of the atom via both laser \cite{Olmschenk07,Duan04,Madsen06,Molmer99,Blatt2008,Eschner03} and microwave fields \cite{Mintert01,Lake15}. This isolation can not be achieved with the use of static fields alone and Paul traps \cite{Paul90} use a combination of radio frequency (rf) fields and static electric fields to confine ions. Another approach for confinement is to use Penning traps which use a combination of static electric fields and magnetic fields \cite{Hasegawa05}.

This article will focus on a selection of the most commonly used rf ion traps or those with unique properties which are currently in use. Photons entangled with quantum memories are excellent carriers of quantum information and may be used for entanglement generation. Protocols for high probability excitation \cite{Duan03} and entanglement-based networking robust to some loss, like a quantum repeater architecture \cite{Briegel98}, offers one path to quantum networking. Ion trap technology is advancing rapidly and the focus of improvement varies from group to group. Since our focus is on quantum networking we compare and assess various ion traps in terms of their optical access, size and scalability.

\section{Networks based on trapped ions and flying qubits}\label{NA}
Significant inroads in information processing have been made in engineering qubits from individual ions (with fidelities $>$99.99\%) and controlling two-ion interactions to form quantum gates (with fidelities $\approx$99.9\%)  \cite{Gaebler16,Ballance16}, however, increasing the number of ions trapped has proven challenging. The promise of quantum computing/information has driven efforts aimed at scaling up the system: trapping very large linear chains, schemes of ion shuttling around zones \cite{Moehring11,Kielpinski2002,Hensinger06} or networking ions using photons \cite{Schug13}\cite{Kurz2014}. Here, the focus on the later approach, whereby a photon entangled with an ion is extracted and sent through free-space or optical fiber to a remote site. Coincidence detection of photons from two separate nodes can then create entanglement between the distant qubits \cite{Briegel98}.  

\begin{figure}[!htp]
\centering
\includegraphics[width=0.75\columnwidth]{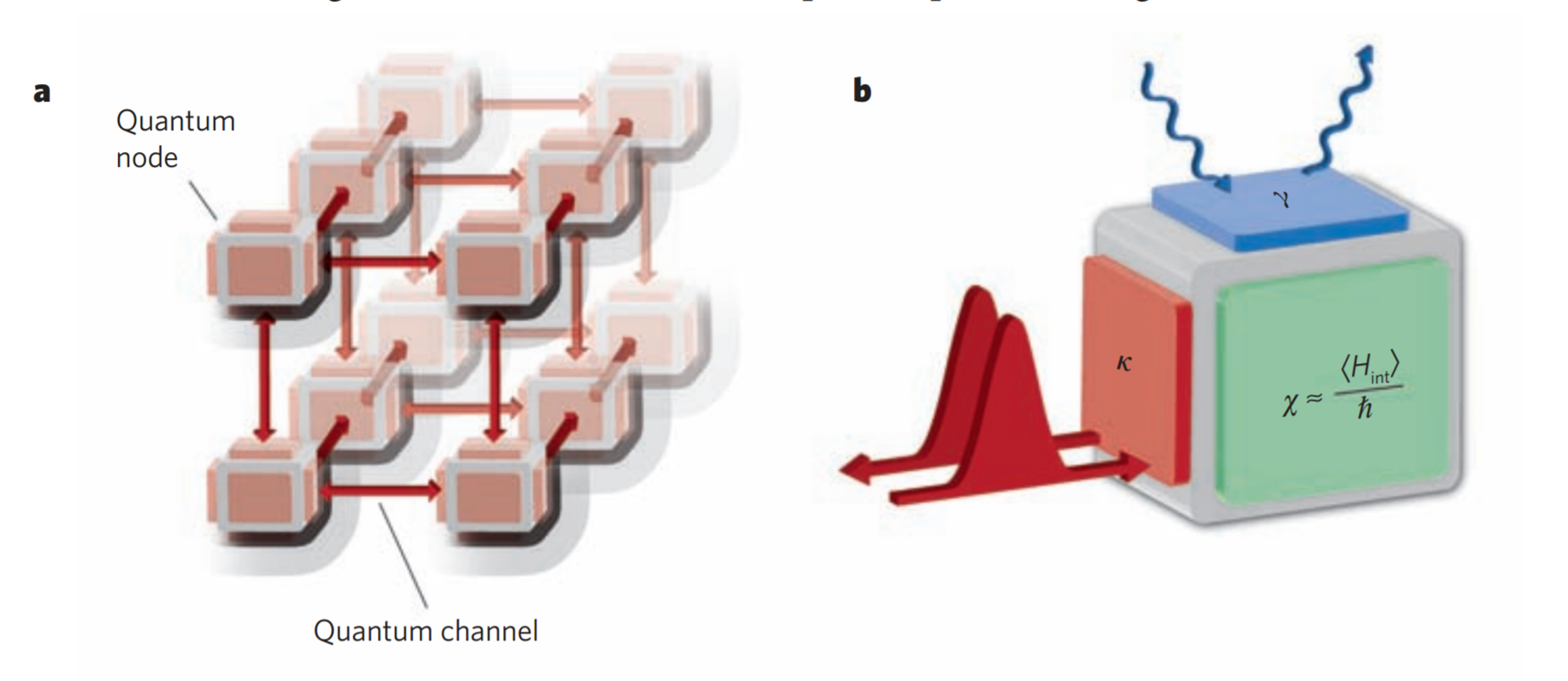}
\caption{Representation of quantum network with elementary building blocks, or nodes, connected via photonic links from \cite{Kimble2008}.}
\label{quantum_internet}
\end{figure}

Fig.~\ref{quantum_internet} shows individual quantum nodes, containing N-qubits, linked via quantum channels \cite{Kimble2008}. Each quantum node contains one or more quantum memories that can emit photons entangled with the memory. Unlike classical networks where amplification may be employed to extend the range between nodes, quantum information cannot be noiselessly amplified \cite{Wootters82}, so instead we use coincident detection of flying qubits between neighbors to propagate information. A quantum repeater scheme \cite{Briegel98} uses nested entanglement protocols to generate entanglement between remotely situated nodes with polynomial overhead time. In this scheme, photons are particularly convenient to transmit information and serve as flying qubits in a network to achieve low-loss and long-range transmission.

\subsection{Photon collection probability}\label{sec:heating}
Remote entanglement and quantum networking protocols require the production, collection and detection of single photons from trapped ions \cite{Cabrillo99,Duan03}. The efficiency with which these photons can be collected has a large impact on the rate at which any entanglement or networking events can be carried out. The probability of detecting a photon entangled with an ion's internal state is given by \cite{Hucul15}.

\begin{equation}
P=P_{dec}Q_{E}T_{f}T_{L}T_{o}\left(\frac{\Omega}{4\pi}\right)
\label{eqn_phot_prob}
\end{equation}

\noindent where $P_{dec}$ is the probability of the ion undergoing the desired decay from an excited state to the qubit state, $Q_E$ is the quantum efficiency of the photon detector, $T_{f}$ is the coupling efficiency into a single-mode fiber, $T_{L}$ is the transmission loss through the fiber, $T_o$ is the optical transmission through any other optics in the photon's path and $\frac{\Omega}{4\pi}$ is the fraction of the solid-angle subtended by the photon collection optics. Equation~\ref{eqn_phot_prob} gives the probability of detecting one photon emitted from one ion. To distribute entanglement amongst $n$ nodes this probability scales to the power of $n$ making it important to maximize this photon detection probability at a site. There is another scheme, which relies on low-probability ion excitation \cite{Cabrillo99}, and in this case entanglement scales linearly in $P$, however with the stringent requirement of optical path length stability. Here, we propose to use the entanglement scheme described in \cite{Duan03} as it is possible to achieve an ion-photon entanglement success probability approaching unity ($P_{dec} \rightarrow 1$), with the path length only requiring to be stable within the decay time of the ion (typically $\sim$10 ns).

Optimizing the photon collection is important for networking nodes with flying qubits as this directly translates into improved entanglement probabilities.  One method for maximizing the detection probability is to utilize trap architectures which allow for the incorporation of high numerical aperture (NA) lenses to collect as many of the emitted photons as possible. Recent work with a custom-designed multi-element high NA lens has significantly improved entanglement rates by an order of magnitude \cite{Hucul15}. However, there is a limit to how much light can be collected in this manner as an ion trap will always have electrodes, at least partially surrounding the ion shadowing it from any collection optics. 

We are interested in maximizing the collection angle with the use of novel asymmetric trap structures which also lend themselves to miniaturization allowing collection optics to be placed nearer the ion. A selection of traps of this type which will be discussed in section \ref{sec:asym_nonfab}. However, even in this improved geometry, there can be undesirable effects of higher NAs such as polarization mixing, which can negatively effect the fidelity of ion-photon entanglement \cite{Streed09,Siverns16}. In practice, it can be convenient to have a combination of lenses including a lower NA lens tailored for ion-trapping/imaging and a higher NA lens tailored for photon collection. High NA lenses, although desirable for photon collection, offer a smaller depth of focus and are more difficult to align compared with lower NA lenses. Optical access is important when assessing traps for their use as a node of a quantum network and, as different trap geometries are presented, we will comment on their ability to allow for use of both an imaging (lower NA) and photon collection (higher NA) optic. Improvements to photon-extraction rates or other performance metrics, requires understanding trap architectures and identifying topologies which offer the best performance for the desired protocol.

Although one could make increasingly higher NA lenses, their size would make impractical integration into a setup, especially considering the community's efforts at trap miniaturization. Alternatively, the optics/lens can be brought closer to the ion or placed inside the vacuum chamber. As trap architectures are miniaturized, the ions may come in close proximity to such surfaces and can succumb to anomalous surface or material potentials. Typical microfabricated ion traps have ion-surface distances in the sub 100 $\mu$m range \cite{Hughes11} (see Table~\ref{tbl:fab_asym_traps}). As the ion height above the surface, $r$, is reduced, the anomalous heating rate has been shown to increase roughly with an $r^{-4}$ scaling \cite{Turchette00,Deslauriers06}. The scaling study \cite{Deslauriers06} was done in a needle-type trap (discussed in Section~\ref{cavities}) which offers a particularly convenient electrode configuration for separating the RF electrodes for observing heating effects. The anomalous heating is thought to be attributable to electric field fluctuations and causes motional heating of trapped ions which can lead to decoherence. Careful processing, even in-situ plasma cleaning, has reduced its effects and provided evidence that the noise is not attributable to thermal Johnson noise \cite{hite2013}. 

The dominant, even contributing, mechanisms behind it remain unclear, although parameters that are thought to be relevant include: ion-electrode distance, material composition, frequency dependence, electrode temperature and unwanted deposition of material on electrodes \cite{Sage15,Allcock2012}. 

%----------------------------------------------------------------------------

\subsection{Photon detection requirements in ion-photon quantum networks}\label{sec:snspd}
One long-standing limit to increasing detection probabilities has been the quantum detection efficiency (QE). Most experiments use either avalanche photodetectors (APDs) or photomultiplier tubes (PMTs) for detection. Generally speaking, APDs have higher QEs than PMTs, although they also have higher dark counts. However, there are now APDs with dark counts in the range of 10 Hz for certain wavelengths (see Table~\ref{tbl:qe}). Such low dark count rates are needed as, in the case of trapped ions, coincidence photon detection rates in two-node networks are in the range of 0.16 s$^{-1}$ \cite{Olmschenk486} and more recently 2.1 s$^{-1}$ \cite{Hucul15}, so given these levels of photon detection, an ideal detector would have a high QE coupled with low dark counts. It should be noted that the detector can be gated and so that only the dark count number in this gate period is required to be much less than one. Although there are outstanding detectors available, these detectors do not perform well at wavelengths emitted by trapped ions and Table~\ref{tbl:qe} presents several ion wavelengths and representative QEs for detection with PMTs and APDs. The QEs represent typical values commercially available for the model with the lowest dark counts. Other single photon detectors, for example APDs, may have better QEs but this is typically at the expense of higher dark counts, for example, for 650 nm, a QE of 73\% is available but it has 100 Hz dark count \cite{excelitas}.  

\begin{table}[ht]
\centering
		\begin{tabular}{|c|c|c|c|c|c|}
		\hline
		Species & Wavelength & \multicolumn{2}{c|}{PMT} & \multicolumn{2}{c|}{APD}\\
		        & (nm) & QE \% & Dark counts [s$^{-1}$] & QE \% & Dark counts [s$^{-1}$] \\
        \hline
        $^{138}$Ba$^+$ & 493 nm & 11 & 15 &  70 & 10  \\
		$^{138}$Ba$^+$ & 650 nm &  20 & 100    & 65 & 10  \\
        $^{171}$Yb$^+$ & 369 nm & 13 & 15 &  -  & - \\
		$^{88}$Sr$^+$  & 422 nm & 15 & 10 & 62 & 10  \\
        \hline
        \end{tabular}
\caption{Representative ion species and quantum efficiencies (QE) of photomultiplier tubes (PMTs) and avalanche photodetectors (APDs). PMTs/APDs with the lowest dark counts are presented. Where the PMT (APD) QE numbers are from datasheet Hamamatsu: 11870-01 (Laser Components: COUNT-10B), with the exception of Ba$^+$ PMT QE at 650 nm which is from datasheet Hamamatsu: H7421-40).}
\label{tbl:qe}
\end{table}

Remarkable progress has been made in superconducting single photon detectors (SNSPDs)\cite{Gaudio16}. These devices operate at cryogenic temperatures, and unlike transition edge sensors \cite{Lamas13}, SNSPDs have excellent recovery times (approximately 100 ns or better and with no after-pulsing, as seen in APDs), high quantum efficiencies ($>$ 90\% for telecom photons) \cite{Marsili13}, and are now commercially available in closed-system cryostats. Although SNSPDs were originally designed for telecom photon detection, where traditional detectors have notoriously low QEs and high dark counts, they can, in principle, perform very well at trapped ion wavelengths. 

\begin{figure}[!htp]
\centering
\includegraphics[width=0.75\columnwidth]{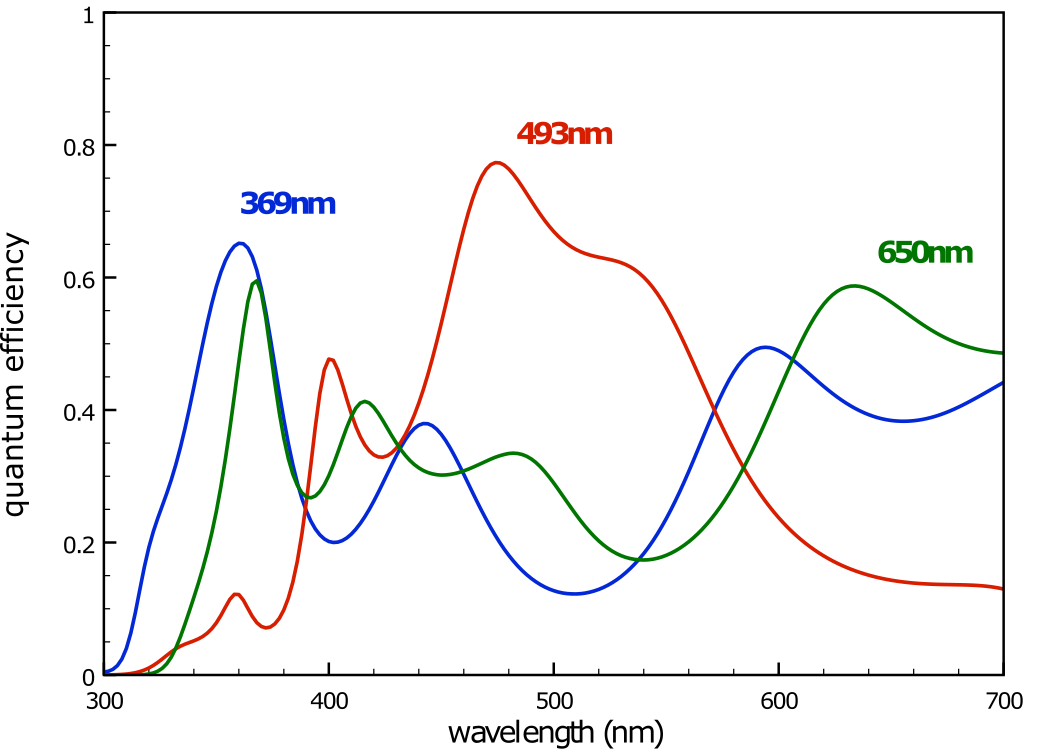}
\caption{Projected normalized quantum efficiency of three tungsten-silicide SNSPDs optimized for detection of Yb$^+$ photons at 369 nm (blue trace) and Ba$^+$ photons at 493 nm (red trace) and 650 nm (green trace), where the expected dark counts are $<$1 Hz for all the traces.}
\label{snspd}
\end{figure}

The fast recovery times without after-pulsing makes them leading contenders for detection devices in quantum systems with a high photon production rate \cite{Rath2015}. This time scale may be much shorter than needed given the rate of emission of photons from trapped ions, however, the exceptionally low dark counts of SNSPDs is particularly suited for photon detection from ions. The uv wavelengths emitted by many trapped ions are, in the case of SNSPDs, actually advantageous as these photons are more energetic and better suited for being measured by the SNSPD as they strongly change the local current density in the superconducting material. SNSPDs can, in principle, be engineered to have high QEs and have essentially zero dark counts ($<$ 10 Hz) \cite{Hadfield2009}. In fact, at such wavelengths, the dark counts from blackbody radiation, the primary noise contributor, is extremely low \cite{Yamashita11}. Our modeling of the tungsten-silicide SNSPDs show we can optimize them for different wavelengths and potentially achieve very good quantum efficiencies (see Fig.~\ref{snspd} \cite{Anant16}). In this model, we calculated the absorption in the tungsten silicide layer using Fresnel equations and an effective index for the thin film containing the patterned nanowire. We plan to test the quantum efficiency and dark counts of these custom detectors which will be housed in available slots in an existing commercial SNSPD cryostat \cite{Siverns16}. These detectors would allow us to detect photons from both ytterbium (369 nm) and barium (493 nm and 650 nm) ions and increase the photon detection probability, given in Eqn.~\ref{eqn_phot_prob}, at one node. Recent results have been reported for in-vacua SNSPD integration with an ion trap \cite{Slichter16}.  

Other factors that affect the photon detection probability shown in Eqn.~\ref{eqn_phot_prob} are the propagation losses in optical fibers at photon wavelengths emitted by trapped ions, $T_L$, and the coupling of a photon into a fiber (current experiments have this in the 20\% range \cite{Hucul15}). Substantial propagation loss can be present even in wavelength-specific optical fibers \cite{Siverns16}. Efforts are currently underway to also improve propagation losses using quantum frequency conversion \cite{Kumar90}\cite{Lenhard:17} and our in-house plan focuses on implementing frequency conversion on barium photons \cite{Siverns16,Feyer16,Li16}. In this approach, the native wavelength emitted by the ion is frequency down-converted into a regime more amenable for fiber transmission, ideally in the telecom regime. For example, 650 nm photons may be converted in a single-stage into the lower O-band range. Also, one can envisage frequency converting 493 nm Ba+ photons into 780 nm photons for hybrid networking with a neutral atom-based memory whereas, a second stage of conversion could result in 1550 nm telecom photons \cite{Siverns16}. Using frequency conversion would give $T_{L} \approx 0.18$ dB/km for telecom wavelengths rather than  typical $T_{L} \approx 50 (15)$ dB/km for 493 (650 nm), a significant improvement even when using relatively short lengths of optical fiber. The conversion can, in principle, occur with relatively high efficiencies and it has already been shown that a typical converted photon's quantum-coherence is well-preserved \cite{Zaske12}. 

Once photons arrive from a remote node, they can be frequency up-converted for entanglement generation with a locally situated trapped ion node to make a two-node network. In this case, having high quantum detection efficiency for both the telecom and native ion wavelength is important, and in principle this could be achievable with several SNSPD detectors mounted in one cryostat head. The projected numbers shown for quantum efficiencies in Fig.~\ref{snspd} \cite{Anant16} are comparable or better than those shown in Table~\ref{tbl:qe} with the additional benefit of a projected $<1 $ Hz dark count. A longer-term vision includes integration of an optical cavity, resonant with both Ba$^+$ (serving as a communication ion with a frequency converted emitted photon) and Yb+ (serving as a memory ion utilizing its hyperfine clock-states), with a microfabricated trap for improved photon collection beyond high NA lenses. 

%------------------------------------------------------------------------
\section{Principles of Radio-frequency (rf) Paul traps}\label{principles}
\begin{figure}[!htp]
\centering
\includegraphics[width=0.3\columnwidth]{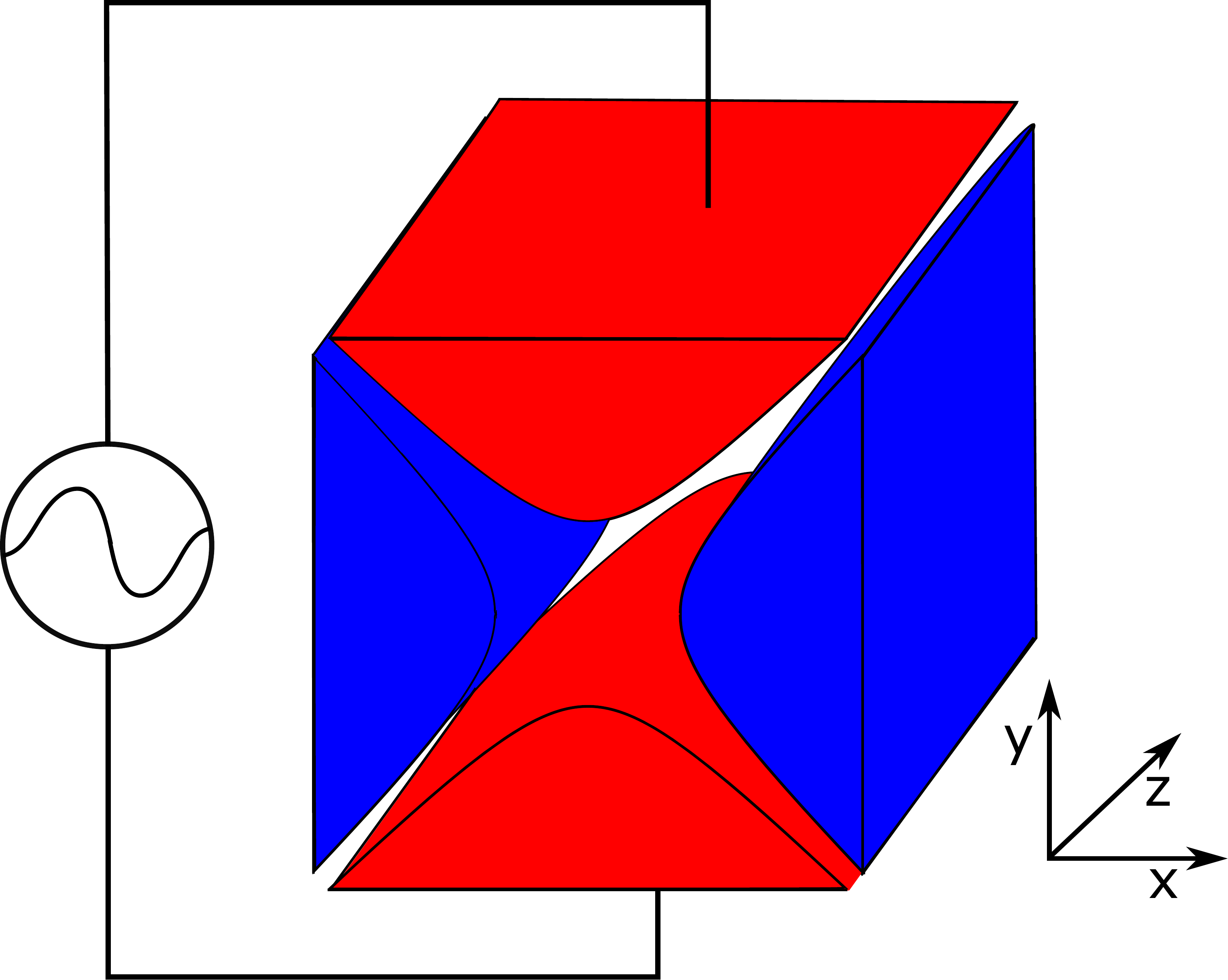}
\caption{Diagram showing an idealized hyperbolic electrode structure for a Paul trap. A trapping potential of the form in Eqn.~\ref{pond_pot} can be created by either applying rf voltages to the red electrodes (as depicted) and static voltages to the blue electrodes or by applying the same rf voltages to all four electrodes but with a $\pi$-phase shift between the blue and red electrodes. To achieve confinement in the z-direction static voltage end-cap electrodes need to be introduced.}
\label{hyperbol_trap}
\end{figure}

To create an electric field capable of confining an ion, an rf Paul trap uses a combination of rf and static voltages applied to electrodes near the ion. An idealized hyperbolic electrode structure for a linear trap of this type is shown in Fig.~\ref{hyperbol_trap}. In this configuration, a quadrupole trapping field is created in the x and y-directions while no trapping field is present along the z-axis. Confinement along the z-axis can be achieved with the addition of two end-cap static voltage electrodes placed at either end of the trap. Although the physical trapping electrodes may vary in their geometry, for example, electrodes fabricated around a fiber tip \cite{Hiroki13}, or the complex electrode arrays found on surface traps \cite{Hughes11}, the confinement we describe here serves as the basic trapping principle for all rf Paul traps. 

For infinitely long electrodes or those sufficiently far from the end caps, the rf potential is uniform in the $\mathbf{z}$ direction and is given by,

\begin{equation}
\Psi(x,y,t)=V(x,y)\cos{\Omega_T t}
\end{equation}

\noindent where $V(x,y)$ is the amplitude of the rf potential and $\Omega_T$ is the applied rf frequency. At time, t, when an ion is placed in this field it is subject to a force towards the center of the trap along one axis but not in the other. Due to the oscillatory nature of the field, a half cycle later this situation is reversed, and the ion undergoes periodic motion in the trap. Near the trap center a perturbative expansion of the electric field gives a non-vanishing time-averaged force on the ion. The pondermotive pseudopotential is then given by \cite{Dehmelt67}

\begin{equation}
\psi(x,y)=\frac{e}{4m\Omega_{T}^2}\left\vert\nabla V(x,y)\right\vert^2
\label{pond_pot}
\end{equation}

\noindent where $e$ is the charge of an electron and $m$ is the mass of the ion. Given infinitely long electrodes along the z-axis, Eqn.~\ref{pond_pot} can be shown to be \cite{Madsen04}

\begin{equation}
\psi_{paul}=\frac{e^2V_0^2\eta^2}{4m\Omega_{T}^2r^4}\left(x^2+y^2\right)
\label{paul_pot}
\end{equation}

\noindent where $V_0$ is the amplitude of the applied rf voltage, $\eta$ is a geometrical efficiency factor between one and zero \cite{Madsen04} and $r$ is the distance from the ion to the nearest electrode. For a perfectly hyperbolic geometry (as shown in Fig.~\ref{hyperbol_trap}) $\eta$ will equal one. The motion of the ion inside this field can be approximated as secular harmonic motion \cite{Wineland98} with frequency

\begin{equation}
\omega_{s}=\frac{eV_0\eta}{\sqrt{2}m\Omega_Tr^2}.
\label{paul_sec}
\end{equation}

The stability region for trapping may be found by looking at the ion's motional dynamics in the potential. An ion trapped in the pseudopotential given by Eqn.~\ref{pond_pot} can have its motional dynamics described using the Mathieu equations \cite{Ghosh},

\begin{equation}
\frac{d^2x}{d(\frac{\Omega_Tt}{2})^2}+(a_x+2q_x\cos{2\frac{\Omega_Tt}{2}})x=0
\label{mathieux}
\end{equation}
\begin{equation}
\frac{d^2y}{d(\frac{\Omega_Tt}{2})^2}-(a_y+2q_y\cos{2\frac{\Omega_Tt}{2}})y=0,
\label{mathieuy}
\end{equation}

\noindent where $a_x$, $a_y$, $q_x$ and $q_y$ are stability parameters given by

\begin{equation}
a_x=-a_y=\frac{4eU}{mr^2\Omega_T^2}
\label{a_param}
\end{equation}
\begin{equation}
q_x=q_y=\frac{2eV_0}{mr^2\Omega_T^2}
\label{q_param}
\end{equation}

\noindent where $U$ is a static voltage offset on the time dependent potential $\Psi(x,y,t)$. Only certain values of $a_i$ and $q_i$ will provide a stable trapping field. In the region where $q_i\ll 1$ and $a_i=0$ ($i=x,y$) the motion of the ion in i-axis is given by

\begin{equation}
i(t)=i_0\cos{\omega_{s}t}\left[1+\frac{q_i}{2}\cos{\Omega_{T}t}\right].
\label{ion_motion}
\end{equation}

The motion of the trapped ion described in Eqn.~\ref{ion_motion} possess two main features. There is a high amplitude component at the secular frequency, $\omega_{s}$, and the second is micromotion which is a low amplitude component at the frequency, $\Omega_{T}$, of the time dependent potential, $\Psi(x,y,t)$, applied to the trap electrodes. 

Another form of micromotion is caused by the ion's displacement from the rf null. To minimize this micromotion the ion's equilibrium position should be at the rf null. Imperfections in the design of trap electrodes, or charge build up on dielectric surfaces leading to stray fields near the ion, are prime causes of increased micromotion. Such stray electric fields displace the ion's equilibrium position from the rf null. It is important that these effects must be considered when designing and building a trap. If care is not taken to reduce micromotion, undesirable effects may occur such as broadening of atomic linewidths \cite{Berkeland98,Wineland98} and the reduced stability of trapping multiple ions \cite{Drewsen98}. For the example of quantum networking with trapped ions, the effects of micromotion can result in photons emitted from ions in separate traps becoming distinguishable putting more stringent bounds on the process of photon mediated entanglement between remote ions \cite{Vittorini14}. The ability to minimize and control micromotion is of great importance in designing a Paul trap with an ability to trap a long chain of ions for quantum computation or simulation \cite{Debnath2016,Schindler2013}.

A single trapped ion group often has multiple trap topologies in use, where one trap may be more suited for one task than another. Considerable expertise in ultra-high vacuum techniques, optics, rf and microwave electronics is required to establish a functioning ion trap. However, with the rapid growth of research groups pursuing physics with trapped ions, dedicated trap development research efforts are now well-established. These ion-trap foundries leverage expertise both from seasoned trapped ion researchers and material scientists for optimized performance and fabrication \cite{Moehring11,Clark14,Tabakov15,Doret12,Shu14,Herold16}. The focus of theses foundries has been on the development of asymmetric microfabricated traps, a selection of which will be described in section \ref{Microfab_traps}. Most traps are operated at room temperature, however, the latest technology now includes cryogenic cooled traps which reduce heating rates and background gas collisions resulting in increased ion lifetimes \cite{Antohi09,Vittorini13,Labaziewicz08,Niedermayr14}. 

The following sections will give a brief introduction to commonly used rf Paul traps and briefly highlight their salient features. There are two main varieties of Paul traps: symmetric and asymmetric. Symmetric Paul traps are geometries in which electrodes are placed in two or more planes symmetrically around the trapping zone. The other geometry type, the asymmetric Paul trap, generally has electrodes situated in one plane. Traps are tailored to meet desired performance benchmarks as defined by individual research aims, including optical access, size, scalability and individual vs multi-ion control. Below we briefly highlight the importance of photon extraction and measurement aimed at quantum networking.

%----------------------------------------------------------------------------
\section{Symmetric Paul traps}
\begin{figure}[!htp]
\centering
\includegraphics[width=0.75\columnwidth]{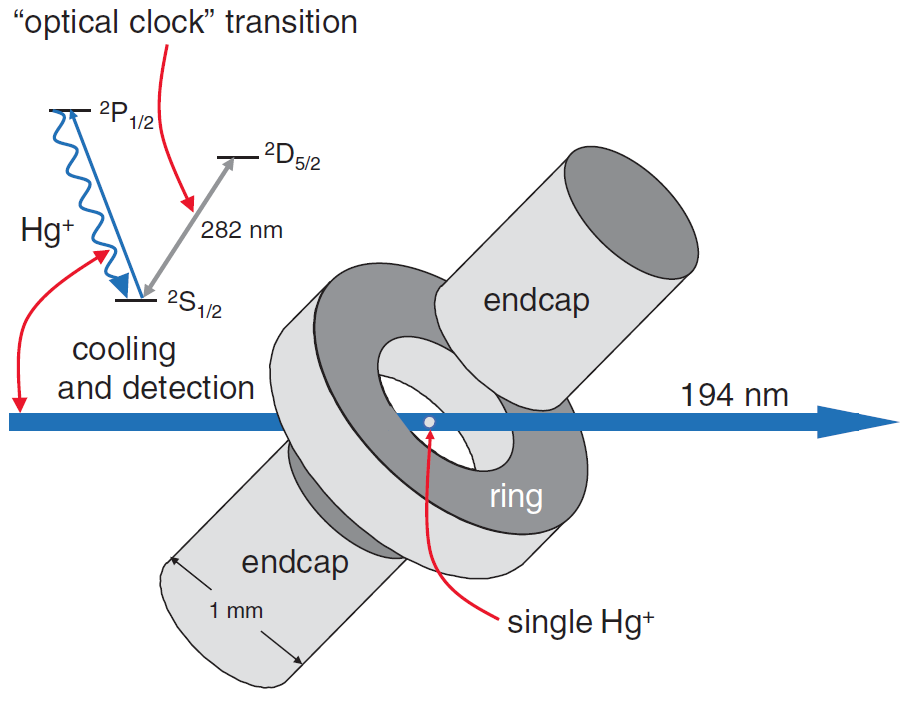}
\caption{Diagram of a ring and end-cap trap used at NIST to carry out trapped ion studies, including precision optical clocks, using a single trapped Hg${^+}$ ion \cite{Bergquist86,Wineland13}. Here an rf potential is applied to the ring electrode and the endcap electrodes are subject to static voltage potentials.}
\label{nistring}
\end{figure}

The two most commonly used types of symmetric Paul traps are the rod-type trap and the blade-type trap. In their most simplest of forms, both of these trap types allow the user to trap a single ion while segmentation of static voltage electrodes allows the user to manipulate the position of individual ions in a chain aligned along one axis \cite{Urabe05}. These traps often have good optical access and require minimal processing. This is in contrast to early traps which, although allowing ground breaking studies into the behavior of single trapped atoms \cite{Bergquist86,Wineland13}, have limited optical access compared with current implementations of the Paul trap. An example of one of these early traps is shown in Fig.~\ref{nistring} which was used to carry out precision optical clock experiments using trapped Hg${^+}$ ions. While providing a suitable trap in which to carry out these challenging experiments the optical access is limited by the placement of the electrodes and so would not lend itself to quantum networking as well as other designs described later in this paper.

\subsection{Symmetric rod traps}
\begin{figure}[!htp]
\centering
\includegraphics[width=1\columnwidth]{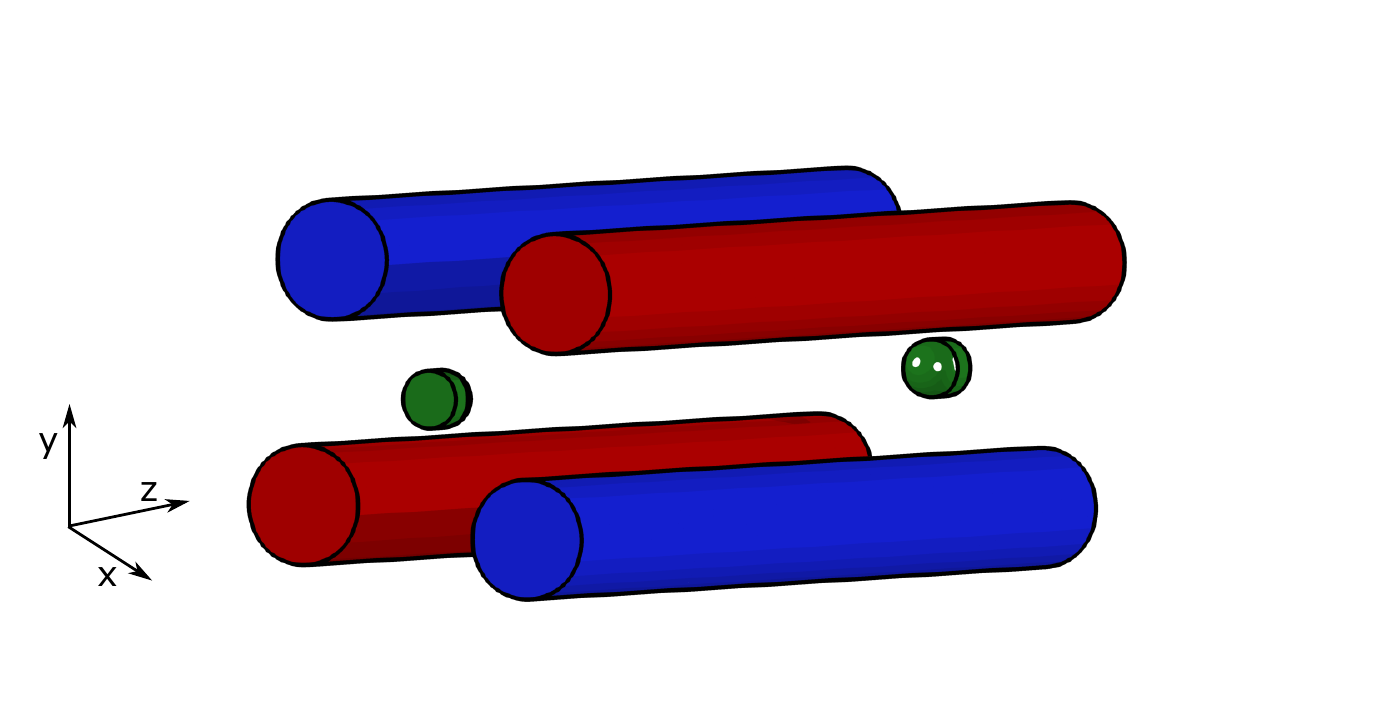}
\caption{A typical four-rod Paul trap. The rf electrodes are shown in red, static voltage electrodes in blue and end-cap electrodes in green.}
\label{four_rod}
\end{figure}

\begin{figure}[!htp]
\centering
\includegraphics[width=1\columnwidth]{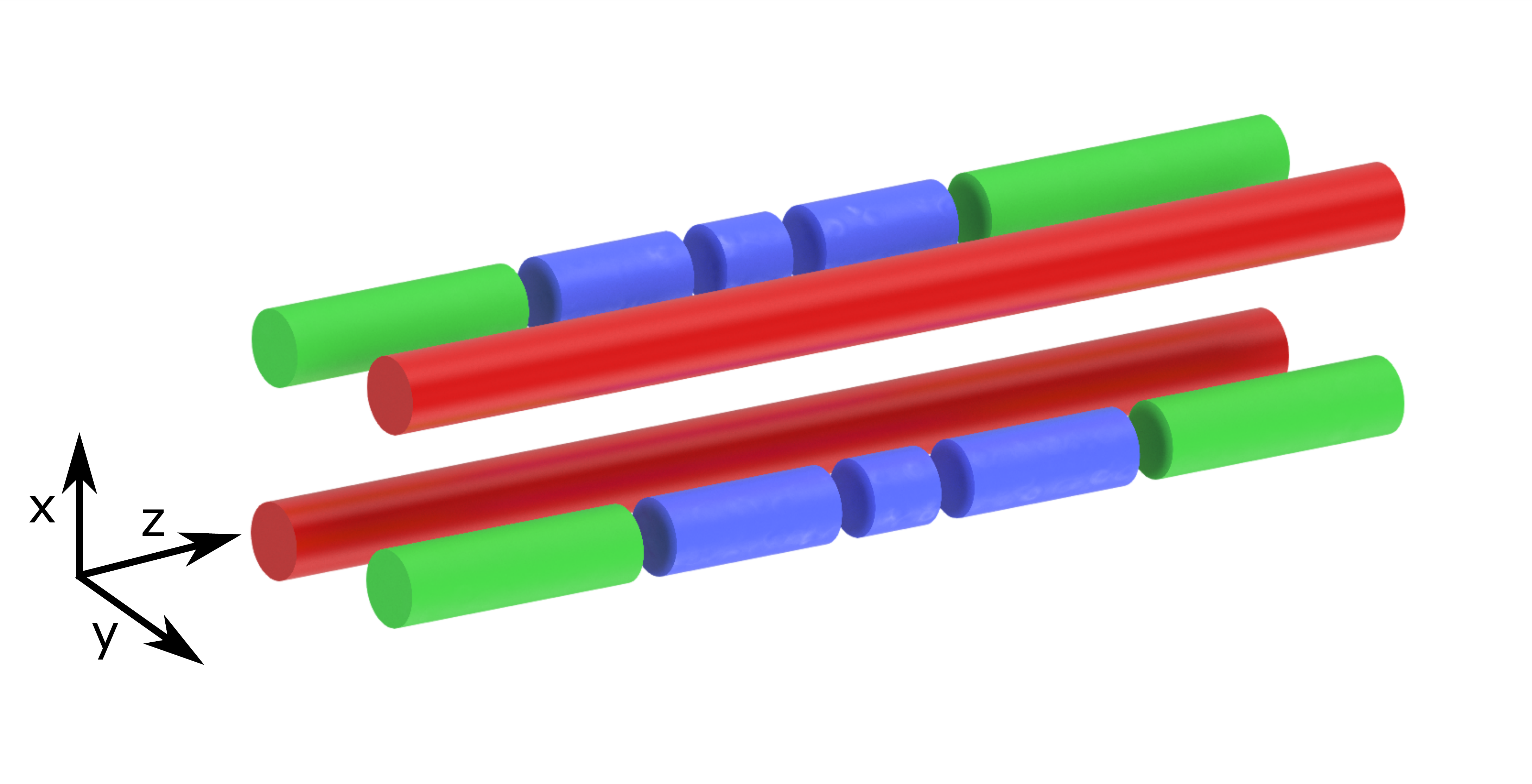}
\caption{A typical four-rod Paul trap with segmented static voltage electrodes. The rf electrodes are shown in red, static voltage electrodes in blue and end-cap electrodes in green. A different static voltage can be applied to each segment allowing manipulation of the trapping potential.}
\label{seg_rod}
\end{figure}

The four rod trap, as shown in Fig.~\ref{four_rod}, is a simple ion trap design. Commonly four rods are used in a two-layer configuration. The electrode configuration shown in Fig.~\ref{four_rod} creates a rf null along the whole of the z-axis allowing ions to be trapped in a linear chain. In order to provide confinement along the z-axis end-cap rods (shown in green) are placed at either extremity of the trap. This comes at the expense of reduced optical access in the trap's axial direction (z direction in Fig.~\ref{four_rod}). However, it is possible to incorporate these end-cap electrodes into the existing static voltage electrodes (shown in blue) by segmentation of these electrodes. A different voltage can then be applied to each segment and the outermost segments can act as end-cap electrodes. An example of segmentation in a rod trap design is shown in Fig.~\ref{seg_rod}.

Typical trap dimensions in the radial (x-y direction) tend to range from 100s of micro-meters to mm resulting in relatively low heating rates \cite{Turchette00} and rf voltages in the 100s of volts range.

\subsection{Symmetric Blade traps}\label{s:blades}
\begin{figure}[!htp]
\centering
\includegraphics[width=1\columnwidth]{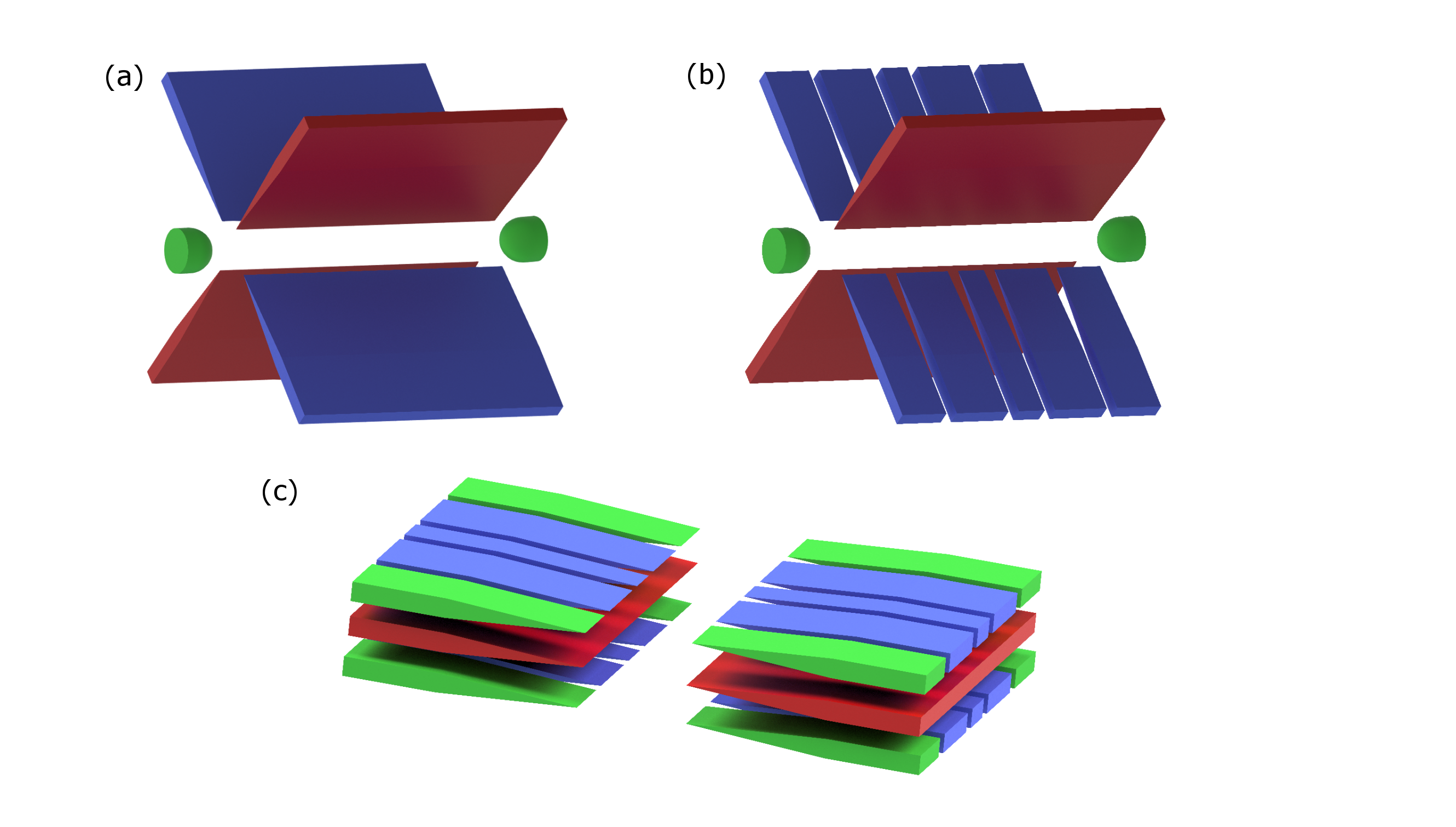}
\caption{Various blade trap designs. All electrodes are colored coded as follows: Static end cap electrodes - green. Static voltage electrodes - blue. Radio-frequency electrodes - red. (a) Shows a basic non-segmented trap. (b) Shows a segmented blade trap where the outer static voltage segments are used to form end cap electrodes allowing for optical access along the trap axis and additional manipulation of the trapping potential. (c) Shows a three-layer variant of a segmented blade trap design.}
\label{seg_blade}
\end{figure}

An alternative design to the rod trap is found by replacing the rods with thin blades. These blades can be placed in such a way as to increase the possible collection angle for photons emitted from the ion and allow for a closer separation, as compared to rods, between the electrodes. Significant experimental results have been achieved in these traps ranging from quantum simulation and computation to fundamental clock physics \cite{Keller16,Islam13,Smith2016,Debnath2016}. A trap of this type's optical access, reconfigurability of trapping potentials and ability to trap multiple ions makes it a good candidate for quantum networking. An example of different variations of blade traps with monolithic blades placed symmetrically around the ion are shown in Fig.~\ref{seg_blade}. This creates a trapping potential only in the radial direction of the geometry. In order to confine the ions along the axis of the trap, end cap static voltage electrodes (shown in green) must be placed at either end of the trap axis. This type of trap is capable of trapping a large chain of ions with radial secular frequencies of up to several MHz, depending on the amount of rf voltage applied.

\begin{figure}[!htp]
\centering
\includegraphics[width=0.75\columnwidth]{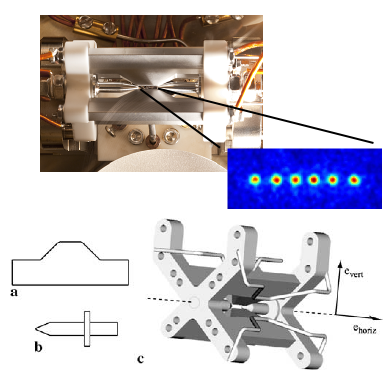}
\caption{The Innsbruck blade trap \cite{SchmidtKaler03}. The top image is a photograph of the assembled trap and superimposed is an inset of a CCD image of a chain of ions trapped along its z-axis. The bottom image shows (a) the shape of the four stainless steel blade electrodes, (b) the molybdenum end-cap electrodes and (c) the arrangement of the electrodes including compensation electrode rods. The trap components are mounted on a Macor holder.}
\label{innsbruck_blade}
\end{figure}

An advantage of the trap shown in Fig. \ref{seg_blade}(a) is its simplicity and relative ease of construction, as it is robust against small misalignment in the axial direction due to the monolithic nature of the blades. An example of a trap of this type is shown in Fig.~\ref{innsbruck_blade} and consists of four stainless steel blade electrodes along with molybdenum end-cap electrodes. This trap has the ability to trap long strings of ions, for example a string to 14 calcium ions \cite{Monz11} allowing for quantum gates between ions to be carried out such as c-NOT gates \cite{Schmidt-Kaler2003} and Toffoli gates \cite{Monz09} as well as the creation of entangled states \cite{Haffner2005}. 

\begin{figure}[!htp]
\centering
\includegraphics[width=0.5\columnwidth]{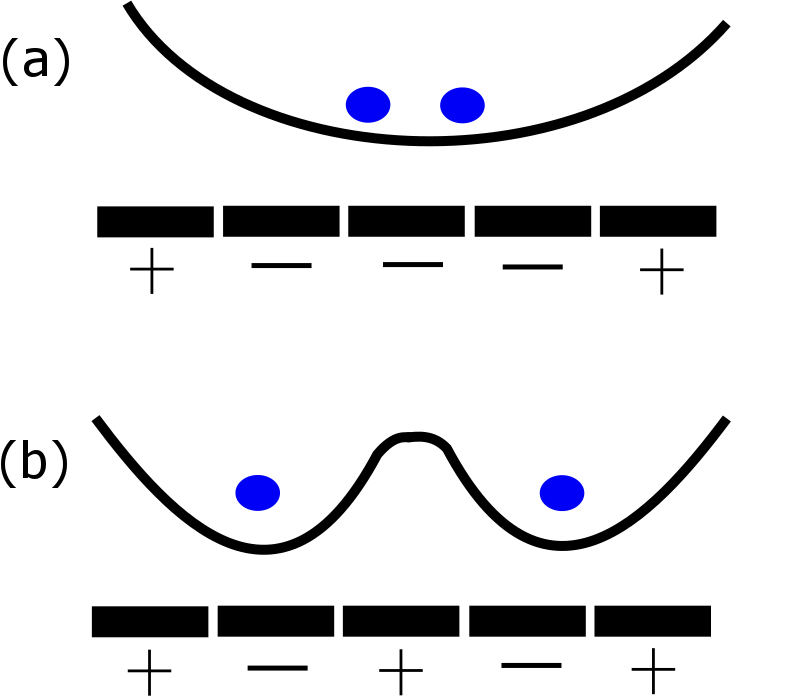}
\caption{Diagram showing: (a) The electric potential created by a set of segmented electrodes and (b) shows how the same set of segmented electrodes can be used to produce two separate trapping wells.}
\label{wedge_hump}
\end{figure}

A disadvantage of non-segmented blade type traps, however, is that control over the ion's axial position (z-axis) is only possible by altering the voltage on the two end cap electrodes. Using this voltage on the endcaps allows for improved micromotion compensation or adjusting the inter-ion spacing when trapping a chain of ions. In order for more complex control of the trapping potential, segmentation of the electrodes is required.  Segmentation allows for shuttling of ions into different potential zones \cite{Rowe02,Chiaverini2004,Hensinger06} or engineering potentials resulting in more equidistant ion-spacing.  For example, segmentation can allow for combining of ions into one potential (Fig.~\ref{wedge_hump}(a)) or separation of ions into multiple potentials (Fig.~\ref{wedge_hump}(b)).

\begin{figure}[!htp]
\centering
\includegraphics[width=0.8\columnwidth]{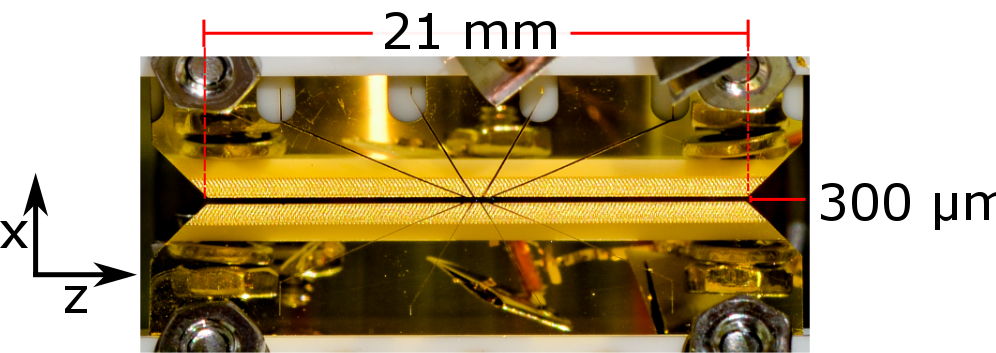}
\caption{Blade trap with segmented static voltage electrodes as used at the University of Maryland [53] and Army Research Laboratory \cite{Siverns16}. The photograph is taken looking along the y-axis (see Fig.~\ref{seg_blade}(b)). The trap contains five segmented static voltage electrodes held in place on a Macor holder. The inner three segmented electrodes are 250 $\mu$m in width and the gaps between the segments are 50 $\mu$m. Both the radio frequency (rf) and static voltage electrodes are constructed from gold coated alumina. Although the upper blade is segmented, unlike the rf blade shown in Fig.~\ref{seg_blade}(b), these electrodes are electrically shorted to one another out of the field of view of the image and one rf voltage is applied.}
\label{JQI_blade}
\end{figure}

The desired ion-control informs the design of the segments and particularly their width (z-direction in Fig.~\ref{seg_blade}(c)). For separating ions, the segment width must be sufficiently small so that ions spaced a few microns apart can experience the field gradient needed to move its position \cite{Nizamani2012}. An example of segmentation is shown in the trap in Fig.~\ref{JQI_blade}. The blades are made from a hard ceramic (alumina) and fabrication techniques are used to sputter gold onto the blades atop a titanium or chromium adhesion layer. The blades are mounted onto a Macor holder for electric isolation of the segments from each other and the vacuum chamber. Relative tilting of the blades in either the x or y direction can prevent micromotion compensation across a chain, nevertheless, significant results have been achieved with blade traps \cite{Keller16,Islam13,Smith2016}.

Another benefit of segmentation is the ability to gain additional optical access. Namely, potentials placed on the electrodes provides z-axis trapping and allows for increased optical access along the z-direction (as seen by comparing Fig.~\ref{seg_blade}(a) to Fig.~\ref{JQI_blade}).  Such a design no longer needs to make use of the traditional end-cap electrodes as the outer segments act as effective end-cap electrodes. 

\begin{figure}[!htp]
\centering
\includegraphics[width=0.5\columnwidth]{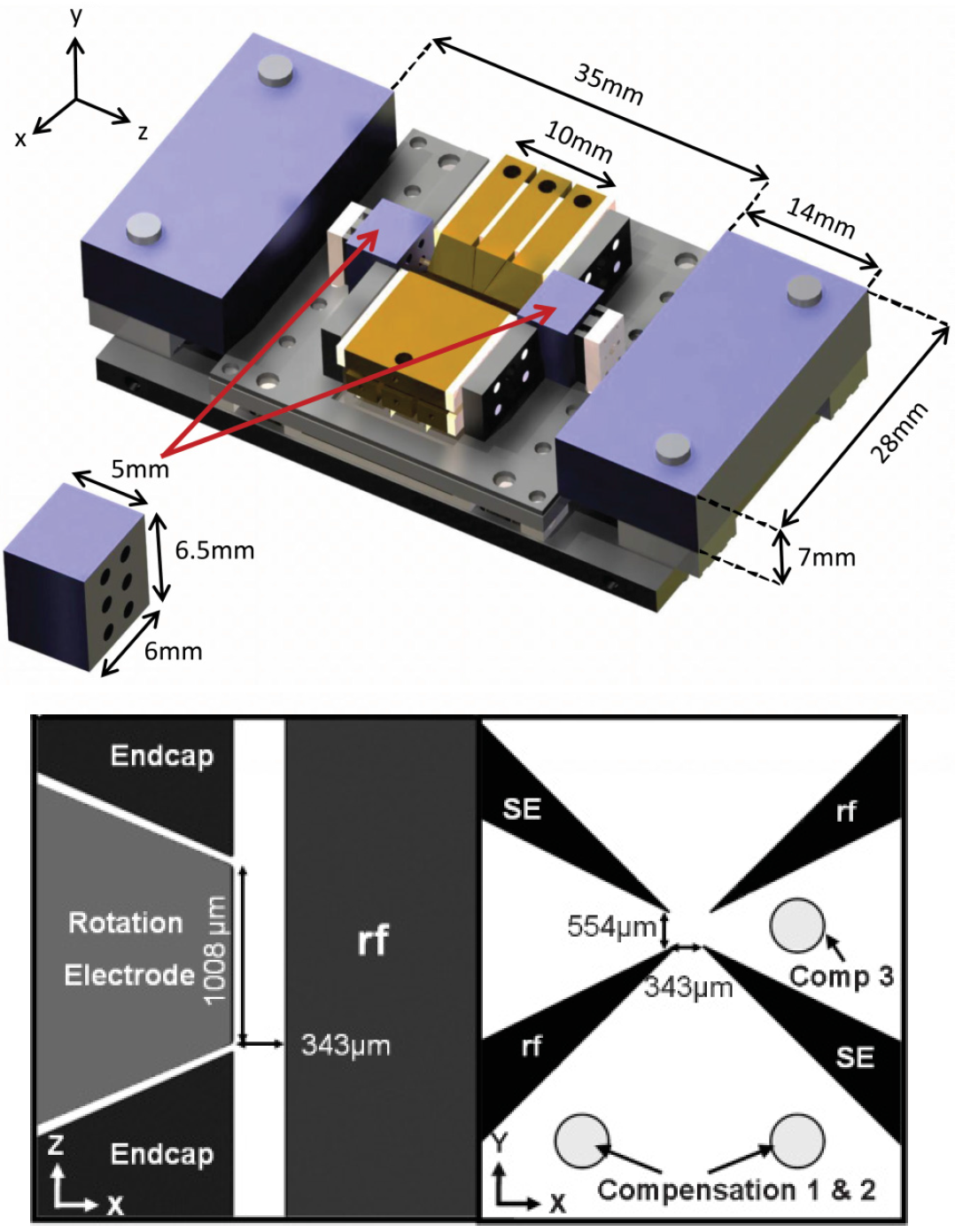}
\caption{Segmented blade trap used at the University of Sussex. (a) Shows an overview of the design containing four samarium cobalt magnets (top diagram shown in blue) have been integrated into the design to provide a magnetic field gradient of around 23 Tm$^{-1}$\cite{Lake15}. (b) Shows the dimensions of the trap which contains three segmented static voltage electrodes held in place on a PEEK holder with a stainless steel cover to reduce exposure of dielectrics to the ion \cite{McLoughlin11}.}
\label{sussex_blade}
\end{figure}

Examples of segmented blade type Paul traps are shown in Fig.~\ref{JQI_blade} and Fig.~\ref{sussex_blade} where the static voltage blade electrode is segmented into five and three electrodes respectively. This leaves the three inner electrodes in the design shown in Fig.~\ref{JQI_blade} which can be used to modify the axial trapping potential in order to more evenly space ions in a string. It is also possible to create two separate trapping regions as shown in Fig.~\ref{wedge_hump}(b). 

\begin{figure}[!htp]
\centering
\includegraphics[width=0.5\columnwidth]{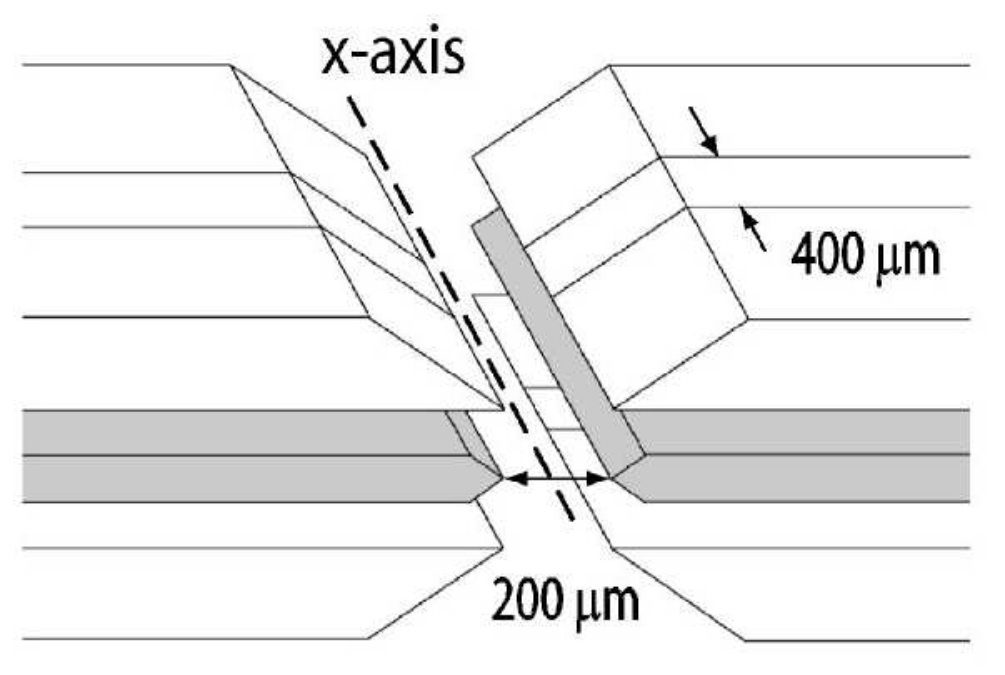}
\caption{A three layer linear ion trap as used, currently, at the University of Maryland \cite{Deslauriers04}. The trap consists of an rf layer (shown in grey) of 125 $\mu$m in thickness and two segmented static electrode layers (shown in white) of 250 $\mu$m in thickness. All layers are made from gold coated alumina and are separated by 125 $\mu$m thick alumina spacers.}
\label{qsim3layer}
\end{figure}

A slight variant of the blade trap is depicted in Fig.~\ref{seg_blade}(c) and consists of three layers of electrodes. The central layer provides the rf voltage and the two outer layers consist of segmented static voltage electrodes. An example of a trap of this type is shown in Fig.~\ref{qsim3layer}. The control of the trapping potential from the segmentation of the static voltage electrodes allows chains of ions to be trapped for studies of quantum simulation of magnetism \cite{Islam13,Korenblit12,Kim2010}, phase transitions \cite{Edwards10,Islam2011} and quantum many-body spin systems \cite{Smith2016,Senko430}. This is a versatile trapping configuration and two layer traps of this type have been fabricated out of PCB material \cite{Pyka2014} and used to perform extremely sensitive optical clock measurements using dual species ion crystals \cite{Keller16}.

For the design shown in Fig.~\ref{sussex_blade}, the segmentation is not sufficient (a minimum of five static voltage electrodes is required) to allow two ion separation but, as traditional end-cap electrodes are not required, it is possible to place magnets at either end of the trap in the axial direction and, therefore, create a strong magnetic field gradient along this axis allowing ground state cooling \cite{Seb15} and quantum gates using long-wavelength radiation \cite{Weidt16}. This would not be possible if the blades were un-segmented and traditional end-caps were used (see Fig.~\ref{four_rod}), as the end-cap electrodes block any access along the axial direction of the trap. However, in some cases, it is possible to engineer these end-cap electrodes to allow optical access in the trap's axial direction for the inclusion of optical fibers for the creation of optical cavities \cite{Brandstatter13}. In Table~\ref{tbl:blade_traps} we summarize salient properties of various blade traps currently in use based on parameters related to those discussed in Section~\ref{principles}.

\begin{table}[ht]
\centering
		\begin{tabular}{|c|c|c|c|c|c|c|}
		\hline
		Trap & RF Drive & RF & Secular & Trap & Ion-electrode & Ion \\
		     & Frequency & Voltage & Frequency & Depth & distance &  Species\\
		          & $\Omega_T/2\pi$  & $V_0$ & $\omega/2\pi$ & $T_D$ & $r$ &  \\
        \hline
        Innsbruck & 23.5 MHz & $\approx$ 1 kV & 1.2 MHz (axial) & several eV  &  800 $\mu$m & Ca$^+$ \\
        Blade \cite{SchmidtKaler03} &&&5.0 MHz (radial)&&&\\
        \hline
        Maryland & $\approx$ 30 MHz & $\approx$ 600 V & 1-2 MHz & $\approx$ 2 eV &  200 $\mu$m & Yb$^+$ \\
        Blade \cite{Hucul15} &&&&&&\\
        \hline
        Sussex & 21.5 MHz & $\lessapprox$ 700 V & 1-2 MHz & $\lessapprox$ 5 eV &  310 $\mu$m & Yb$^+$ \\
        Blade \cite{McLoughlin11} &&&&&&\\
        \hline
        Maryland & 21.5 MHz & $\lessapprox$ 400 V & 0.4-4 MHz (axial) & several eV &  100 $\mu$m &  Cd$^+$/Yb$^+$ \\
        3-layer \cite{Deslauriers04} &&&8 MHz (radial)&&&\\
        \hline
        \end{tabular}
\caption{Summary of typical operating specifications of the blade traps described in the text. These traps offer a relatively straightforward assembly process while having substantial versatility.}\label{tbl:blade_traps}
\end{table}

%----------------------------------------------------------------------------
\subsection{Symmetric traps with integrated optical cavities}\label{cavities}
One method of increasing the photon collection probability is to increase the fraction of the solid-angle that the photon collection optics subtend, as discussed in Section~\ref{NA}. As trapped ions emit photons in all directions, increasing the solid angle of collection can necessitate the use of large collection optics. These optics can become bulky and expensive and are not conducive to scalability, although there has been recent work towards increasing the scalability of photon collection optics using microfabircated Fresnel lenses \cite{Streed11}.

A possible solution to the bulky photon collection optics used in current remote entanglement experiments \cite{Hucul15} is to place the ion inside an optical cavity resulting in a coupling between the ion and cavity mode \cite{Casabone15,Brandstatter13}. The ion will then preferentially emit photons into the cavity mode and collection can then be carried out along the cavity axis. This method can increase the overall collection efficiency \cite{Vogell17} from the $\approx10\%$ \cite{Hucul15} achieved thus far with bulk optics. The optical cavity can increase both $T_f$ and $\Omega$ in Eqn.~\ref{eqn_phot_prob}. However, placing dielectric mirror surfaces close to an ion can result in hard to predict stray electric fields from the mirror surface interacting with the ion. This can make ion-cavity traps extremely difficult to design and operate, as either the mode volume needs to be large or the ion must be rigorously shielded from stray fields \cite{Sterk12}. 

\begin{figure}
\centering
\includegraphics[width=0.9\columnwidth]{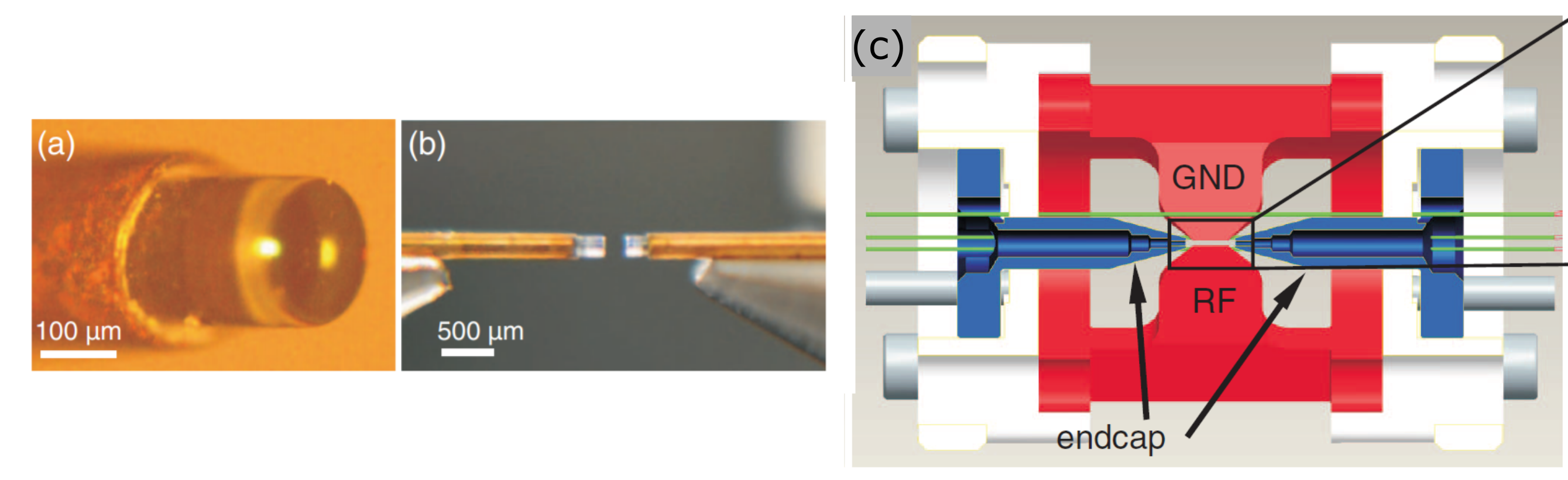}
\caption{Images of the cavity used at the University of Innsbruck \cite{Brandstatter13} (a) Photo of the polished end of the optical fiber  (b) photo of fiber-based Fabry Perot cavity. For vacuum compatibility the fibers are copper coated. (c) A linear rf blade trap with integrated Fabry-Perot fiber cavity, where the ground and rf blade electrodes (in red) and the end cap electrodes (in blue) are indicated. The fibers used to create the cavity are placed into/out of the page.}
\label{innsbruck_cavity}
\end{figure}

\begin{figure}
\centering
\includegraphics[width=0.75\columnwidth]{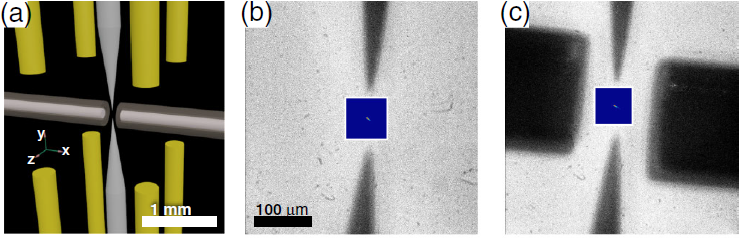}
\caption{A symmetric rod trap used at the University of Bonn \cite{Steiner13}. The rod trap, shown in (a), is created by two rf electrodes shown in grey and is surrounded by static voltage compensation electrodes, shown in yellow. The optical fibers forming the cavity are oriented horizontally. (b) shows the rf rods with a CCD image of an ion superimposed in between the rods. (c) shows the placement of the rf and cavity fibers around the trap.}
\label{bonn_cavity}
\end{figure}

\begin{figure}
\centering
\includegraphics[width=0.35\columnwidth]{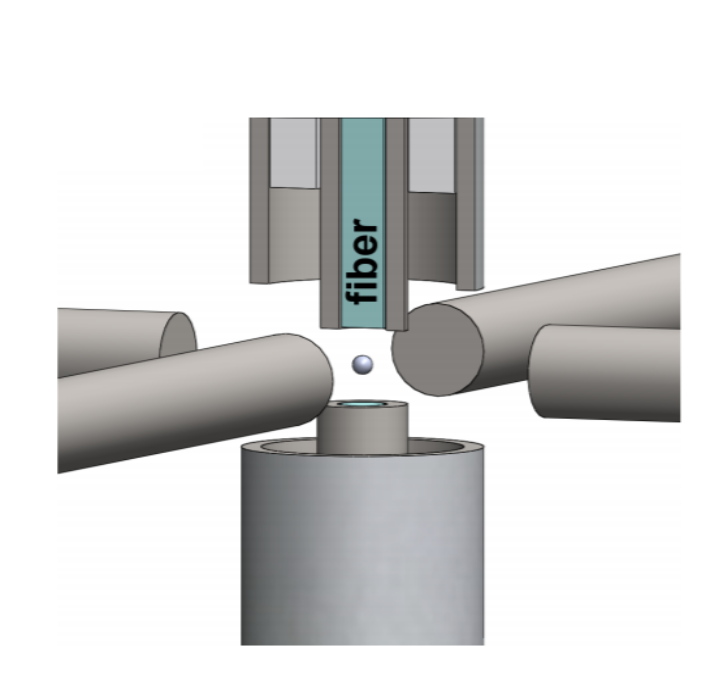}
\caption{An image of the fiber cavity trapped used at Sussex University \cite{Begley16}. The fibers are situated inside stainless steel electrodes and these electrodes, along with two of the four radial electrodes, are used for stray field compensation. Trapping rf voltages are applied to the larger diameter electrodes which are recessed from the end of the fiber.}
\label{sussex_fiber_cavity}
\end{figure}

The trap designs in Fig.~\ref{innsbruck_cavity}, Fig.~\ref{bonn_cavity} and Fig.~\ref{sussex_fiber_cavity} use polished fibers to create an optical cavity centered around the ion. These fibers are polished for the desired curvature (see Fig.~\ref{innsbruck_cavity}(a) for an example of a polished fiber tip) and coated with a highly reflective mirror coating. One advantage of using fibers to create a cavity around a trapped ion is they can allow a substantial reduction in the cavity mode volume. However, care must be taken to avoid charging on the fiber's surface as stray fields can perturb the trap potential resulting in increased micromotion or, potentially, an inability to trap. Fig.~\ref{innsbruck_cavity}(c) shows a bulk blade trap where the fibers are positioned into/out of the page and between opposite blades to minimize the cavity mode volume. In vacuo translation of the this fiber assembly allows for a range of cavity lengths to be tested, as shown in Table~\ref{tbl:cavity_parameters}. Based on cavity finesse measurements very high ion-cavity couplings can be predicated, although these are yet to be verified experimentally at the time of writing.

The cavities shown in Fig.~\ref{bonn_cavity} \cite{Steiner13} and Fig.~\ref{sussex_fiber_cavity} \cite{Takahashi17} are integrated into needle type traps. These cavities are is designed for optical coupling on the P$_{1/2}$ to D$_{3/2}$ transition at 935 nm in $^{171}$Yb$^+$ and the P$_{1/2}$ to D$_{3/2}$ transition at 866 nm in $^{40}$Ca$^+$ respectively. By creating an ion-cavity coupling to these transitions the ions in question can be made to emit photons at wavelengths much more amenable to networking compared to their more commonly emitted S to P transition UV photons at 369 nm and 397 nm, respectively. These longer wavelength photons, say at 935 nm, may be used for hybrid quantum networking with solid-state quantum memories \cite{Meyer15}. Also, the longer wavelength photons also allow for frequency conversion into the telecom band using just one conversion stage, whereas, the UV photons may require two or three stages, substantially reducing the overall efficiency.

To minimize the effect of surface charging on the mirrors the optical fibers in Fig.~\ref{bonn_cavity} \cite{Steiner13} can be pulled back during ion loading when there is a flux of neutral atoms present. The fibers in Fig.~\ref{sussex_fiber_cavity}
achieve increased shielding, compared to bare fibers, by being physically integrated into the trapping structure itself and recessed inside stainless steel tube electrodes. For residual field compensation, voltages are applied to two of four radially positioned rods as well as the inner end cap electrodes, all of which are shown in Fig.~\ref{sussex_fiber_cavity}. At the
time of writing, this setup has achieved the highest ion-cavity coupling factor of $g=5.3(1)$ MHz \cite{Takahashi17}. The optical cavity designs presented here allow cavity lengths, $L$, in the 100s of $\mu$m range and cavity mode waists, $\omega_0$, of several $\mu$m in order to increase the coherent coupling strength, $g$, between the ion and the cavity which scales as $g\propto \omega_0^{-1}L^{-\frac{1}{2}}$ \cite{Brandstatter13}. An additional benefit of trapping a chain of ions is the Coulomb interaction can be used as a communication bus within the chain to allow a large range of quantum information and quantum simulation applications to be studied \cite{Monroe14,Wright2016}.

\begin{figure}
\centering
\includegraphics[width=0.75\columnwidth]{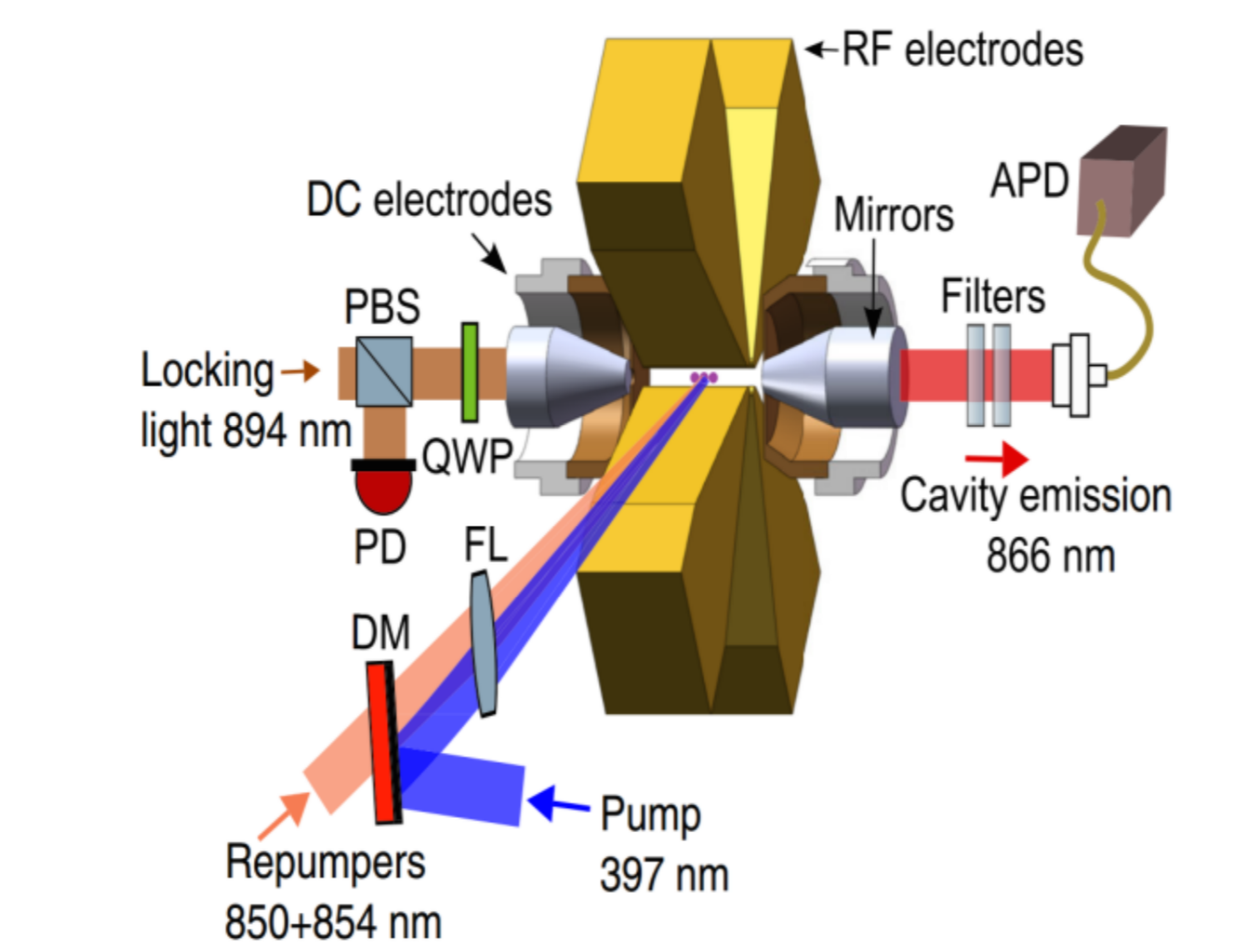}
\caption{The optical cavity trap as used at Sussex University \cite{Begley16}.  This trap uses blade shaped electrodes, as discussed in Section \ref{s:blades}. The cavity is placed along the trapping axis and hence allows for a chain of ions to interact with the optical cavity mode. }
\label{sussex_cavity}
\end{figure}

Another design for trapping ions inside the mode of a cavity is shown in Fig.~\ref{sussex_cavity}. Here, bulk mirrors are used to form an optical cavity along the axial direction of the blade trap. This allows for a chain of ions to be trapped in the cavity's optical mode. A cavity can be aligned with either one ion or a group of ions and when aligned with a group this increases the coupling by a factor of $\sqrt{N}$, where $N$ is the number of ions \cite{Fogarty15}\cite{Begley16}. The traps discussed here that incorporate optical cavities offer one path to extracting photons which are mode-matched well into optical fibers \cite{Hiroki13}, although desired cavity parameters would vary based on the application \cite{Vogell17}. For quantum networking we must balance the need for photon extraction against ion-cavity coupling. The potential increase in photon collection and its known directionality of emission improve the likelihood of long-distance photon transmission for quantum networking when coupled with quantum frequency conversion as mentioned in Sec.~\ref{sec:snspd}. Table \ref{tbl:cavity_traps} shows the salient characteristics and Table \ref{tbl:cavity_parameters} summarizes the cavity parameters of the three cavity traps discussed in this section.

\begin{table}[ht]
\centering
		\begin{tabular}{|c|c|c|c|c|c|c|}
		\hline
		Trap & RF Drive & RF & Secular & Trap & Ion-electrode & Ion \\
		     & Frequency & Voltage & Frequency & Depth & distance &  Species\\
		          & $\Omega_T/2\pi$  & $V_0$ & $\omega/2\pi$ & $T_D$ & $r$ &  \\
        \hline
        Innsbruck & 35 MHz & 130 V & 9.6, 9.9 MHz (radial)$^{i}$ & 1.3 eV$^{i}$ &  170 $\mu$m & Ca$^+$ \\
        fiber cavity \cite{Brandstatter13} &&& 1.9 MHz (axial)$^{i}$&&&\\
        \hline
				Bonn & 22 MHz & 160 V & 2.1, 2.7 MHz (radial) & 550 K &  50 $\mu$m & Yb$^+$ \\
        cavity \cite{Steiner13} &&&1.3 MHz (axial)&&&\\
				\hline
				Sussex & 19.6 MHz & NA & 3.46 MHz (axial) & 0.9 eV &  175 $\mu$m & Ca$^+$ \\
        fiber cavity \cite{Takahashi17}  &&& 1.96 MHz (radial) &&&\\
        \hline
        Sussex & 16 MHz & NA & 1.23 MHz (radial) & $\approx$6 eV &  475 $\mu$m & Ca$^+$ \\
        blade cavity \cite{Begley16} &&&  400-620 kHz (axial)    &&&\\
        \hline
        \end{tabular}
\caption{Summary of typical operating specifications of the Paul traps with integrated cavities described in the text. NA (not available). $^{i}$Based on trap simulations not experimental data.  }\label{tbl:cavity_traps}
\end{table}

\begin{table}[ht]
\centering
		\begin{tabular}{|c|c|c|c|c|c|c|}
		\hline
		Trap & ion number & Cavity length   & g$/2\pi$ & $\kappa/2\pi$ & $\gamma/2\pi$ & C \\
		     & & [mm] & MHz & MHz & MHz &   \\
		                 \hline
        Innsbruck & 1 &  0.131, 0.206 & 31$^{a}$, 41$^{a}$ & 8$^{a}$, 9$^{a}$ & 11.2 & 4.8$^{a}$, 9.3$^{a}$ \\
        fiber cavity \cite{Brandstatter13} &&&&& &\\
        \hline
				 Bonn & 1 & 0.230 &  3.4(2) & 320 & 2 & 0.05 \\
        cavity \cite{Steiner13} &&&&&& \\
				\hline
				Sussex & 1 & 0.367 & 5.3  & 4.2 & 10.3 & 0.30 \\
        fiber cavity \cite{Takahashi17} &&&&& & \\
        \hline
        Sussex & 5 & 5.3 & 0.988 & 0.235 & 22.3 & 0.78 \\
       blade cavity \cite{Begley16} &&&&& & \\
        \hline
        \end{tabular}
\caption{Summary of cavity parameters where the cooperativity $C=(\sqrt{N} g^2)/(2\kappa\gamma)$, where $ N $ is the number of ions. Note that the Sussex blade cavity trap \cite{Begley16} has the optical cavity aligned along the trap axis and hence allows for a string of ions to be coupled to the cavity. $^{a}$Values are calculated values based on mirror finesse measurements and not experimentally measured values.}\label{tbl:cavity_parameters}
\end{table}

%-------------------------------------------------------------------------
\section{Asymmetric rf Paul traps}\label{sec:asym_traps}
Asymmetric Paul traps are geometries which contain electrodes placed on one single plane. The resulting trapping potential is located above the surface of the electrodes.  Asymmetric ion traps are often built using microfabrication techniques and such traps are a leading technology as a scalable ion trap platform \cite{Hughes11}. Scalability will be required in order to build large scale quantum computing, communication and simulation architectures \cite{Monroe14}. Asymmetric geometries can also be built using bulk components and some novel bulk asymmetric trap designs will be described in this paper along with a selection of leading microfabricated geometries. From a networking perspective it may be challenging to incorporate both an imaging diagnostic optic and a photon collection optic in some asymmetric trap configurations. However, there has been dedicated work (as discussed below) aimed explicitly at enhanced photon collection using bulk and/or microfabricated optics.  

%------------------------------------------------------------------------- 
\subsection{Non-microfabricated asymmetric Paul traps}\label{sec:asym_nonfab}
Non-microfabricated traps are advantageous in that they can be cost-effective and require less technical fabrication expertise in material processing. This section will outline a few current bulk asymmetric Paul traps. 

\begin{figure}[!htp]
\centering
\includegraphics[width=0.75\columnwidth]{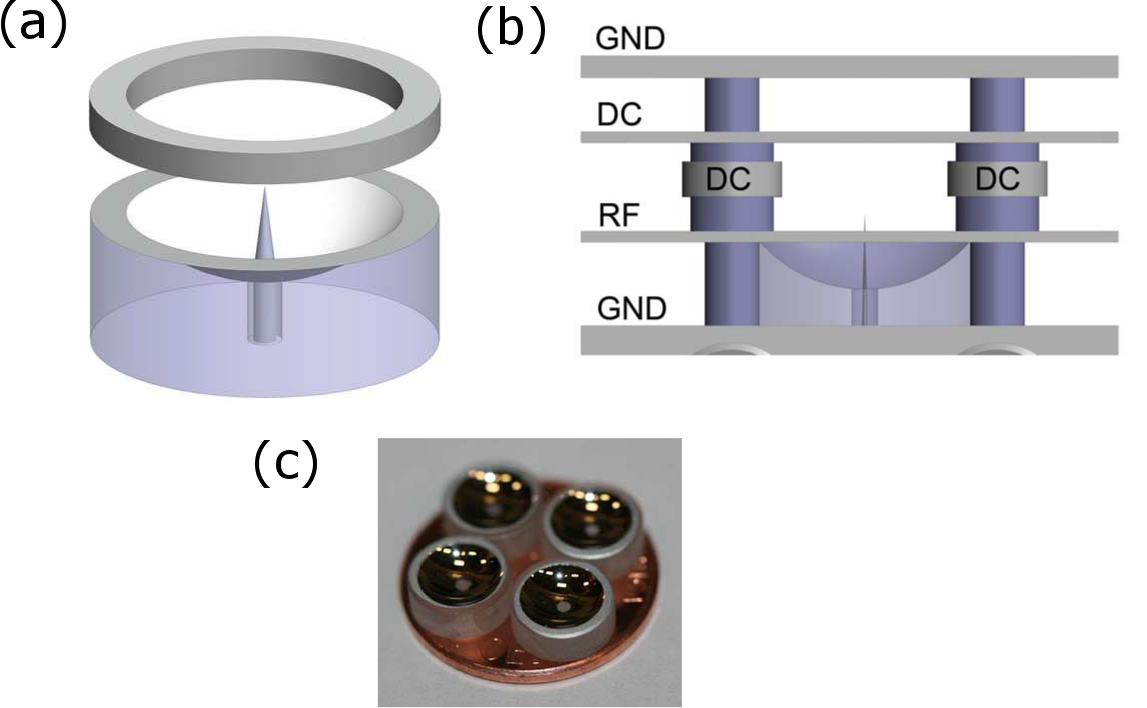}
\caption{The ``tack'' trap used at Washington University \cite{Shu11}. (a) trap drawing showing the curved mirror, where the needle-shaped electrode serves as RF ground and the elevated ring serves as DC ground. (b) profile view of trap illustrating the full assembly and identifying electrodes with RF/DC voltages and (c) four mirrors, one of which is used to construct the trap and for scale, the mirrors have been placed on a US penny.}
\label{washington_tack}
\end{figure}

Asymmetric traps lend themselves very well to increasing the solid-angle of collection of emitted photons compared to symmetric traps. The trap in Fig.~\ref{washington_tack} provides an alternative method for increasing the solid-angle of photon collection for quantum networking compared with the bulk optics in symmetric traps and integrated cavities discussed in previous sections. Here, a needle-type trap is designed around a small aluminum concave mirror in such a way that the ion is trapped at the focus of the mirror. The mirror is made from a conductive material to avoid any charge build up from dielectric surfaces near the ion. The light reflected off of the mirror is collimated and can be focused into fibers for long distance propagation. The mirror collects a larger fraction (25\%) \cite{Shu11} of light compared with the $\approx 10\%$ collected by ex vacuo high NA lens \cite{Hucul15}, and this is desirable for increasing rates of two-photon entanglement protocols (proportional to the square of Eqn.~\ref{eqn_phot_prob} \cite{Duan03}). However generally, needle-type traps are unable to trap an ion chain so it is not readily possible to perform large-scale local quantum information processing or simulation. Yet, this technology could be adapted into a more linear geometry at the expense of some collection efficiency in order to perform some quantum information processing and simulation locally. With a trap of this type, around 35\% of the solid angle of collection can be subtended \cite{Shu11} (although mirror imperfections diminishes the actual collection to $\approx 25\%$) while avoiding the trap assembly complexities and high-performance mirror requirements of optical-cavity based experiments. A similar trap using a parabolic mirror integrated around a needle type trap has also been demonstrated in which 54.8\% of the light emitted from a single ion was collected \cite{Maiwald12}.

\begin{figure}[!htp]
\centering
\includegraphics[width=0.75\columnwidth]{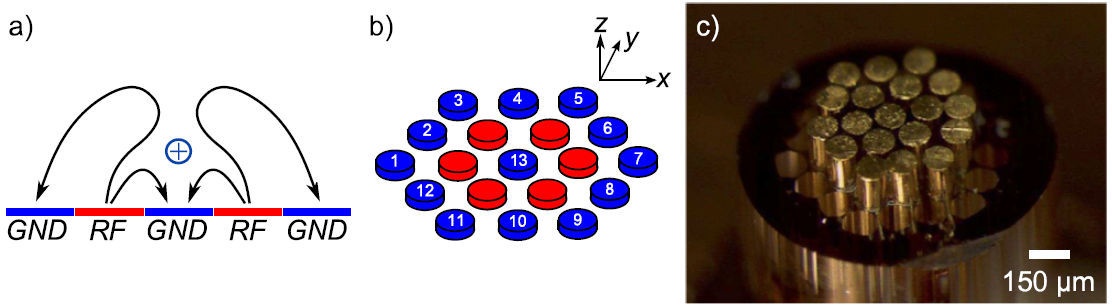}
\caption{An asymmetric trap designed and built at ETH Zurich using photonic-fiber-crystal technology \cite{Lindenfelser15}. (a) shows a sketch of the trapping potential created by the trap electrodes. The ion is trapped at the location marked by the cross. (b) shows the electrode structure with ground electrodes in blue and rf electrodes in red. (c) shows a gold wire photonic-crystal-fiber used to create the trapping structure in (b).}
\label{zurich_pf_trap}
\end{figure}

Figure~\ref{zurich_pf_trap} shows an asymmetric Paul trap created using a gold wire filled photonic-crystal-fiber cane. The fibers were prefabricated into a microstructured regular array using well-established fiber-pulling techniques. The gold rods were inserted into each fiber and extend 2 mm beyond the fiber. They were then polished down to provide a smooth surface protruding 50 $\mu$m from the fiber, as seen in Fig.~\ref{zurich_pf_trap}(c). A trapping potential is formed above the surface of the array, as seen in Fig.~\ref{zurich_pf_trap} (a), when rf and static voltages are applied to the electrodes in the manner shown in Fig.~\ref{zurich_pf_trap} (b). Traps with this lattice structure are capable of trapping 2-dimensional planar arrays of ions in a manner which could be suitable for quantum simulations \cite{Kumph16}. However, as with any trap made from bulk components, the scalability of such devices has limits and considerable efforts have been made to miniaturize both the trap and attendant electronics (such as resistors and capacitors for rf filtering and phase-shift corrections) using microfabrication techniques \cite{Siverns12,Kumph16,Sterling14}. This type of trap could serve as a local computational node with many ions which is an important resource for a quantum network, albeit further development would be needed to couple this 2D system to communication ion(s). Table \ref{tbl:nonfab_asym_traps} gives a summary of these two traps and their salient characteristics.

\begin{table}[ht]
\centering
		\begin{tabular}{|c|c|c|c|c|c|c|}
		\hline
		Trap & RF Drive & RF & Secular & Trap & Ion-electrode & Ion \\
		     & Frequency & Voltage & Frequency & Depth & distance &  Species\\
		          & $\Omega_T/2\pi$  & $V_0$ & $\omega/2\pi$ & $T_D$ & $r$ &  \\
        \hline
        Washington & 23 MHz & 270 V & 420 kHz (axial) & 0.02 eV &  541 $\mu$m & Ba$^+$ \\
        ``Tack" \cite{Shu11} &&& 200 kHz (radial)&&&\\
        \hline
        Zurich & 40.3 MHz & 70 V & 1.8 \& 2.0 MHz  (axial) & 0.1 eV &  88 $\mu$m & Ca$^+$ \\
        Fiber Crystal \cite{Lindenfelser15} &&& 3.8 MHz (radial) &&&\\
        \hline
        \end{tabular}\label{tbl:nonfab_asym_traps}
\caption{Summary of typical operating specifications of the non-microfabricated asymmetric Paul traps described in the text.}
\end{table}

%-------------------------------------------------------------------------
\subsection{Microfabricated asymmetric traps} \label{Microfab_traps}
An advantage of using microfabrication for ion traps is the ability to fabricate small-scale multi-electrode structures, embed the necessary filtering components in the chip, integrate optical elements in close proximity to the trap and allow for improved control over trapping potentials. There has been substantial effort over the last decade in increasing the complexity and scalability of ion trap geometries since the fabrication of the first microfabricated ion trap \cite{Stick05}. For networking, these type of traps can pose a challenge if there is only optical access from one direction as this would inhibit placement of both a low NA lens (for day-to-day operation) and a more alignment sensitive high NA lens (for optimal photon extraction). Although, as we mention below, effort in integrated optics or slotted traps may allay some of these concerns.  

\begin{figure}[!htp]
\centering
\includegraphics[width=0.5\columnwidth]{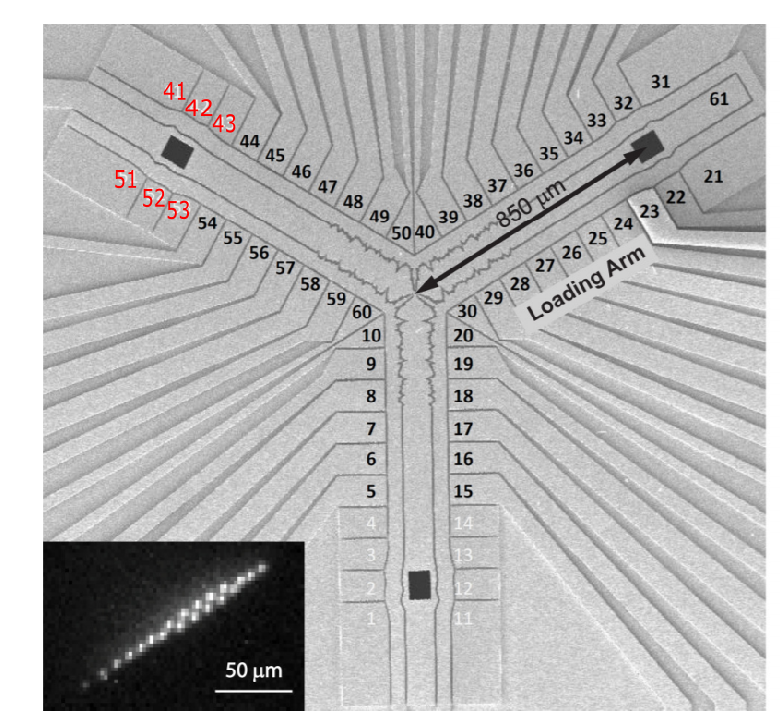}
\caption{The optimized Y-junction surface trap used at the Georgia Institute of Technology and manufactured at Sandia \cite{Shu14}. Black numbered electrodes are connected to digital to analogue converters (DACs) for controlling shuttling voltages. Red numbered electrodes are grounded. The inset shows a chain of ions trapped above the surface of the device.}
\label{georgia_y}
\end{figure}

\begin{figure}[!htp]
\centering
\includegraphics[width=0.6\columnwidth]{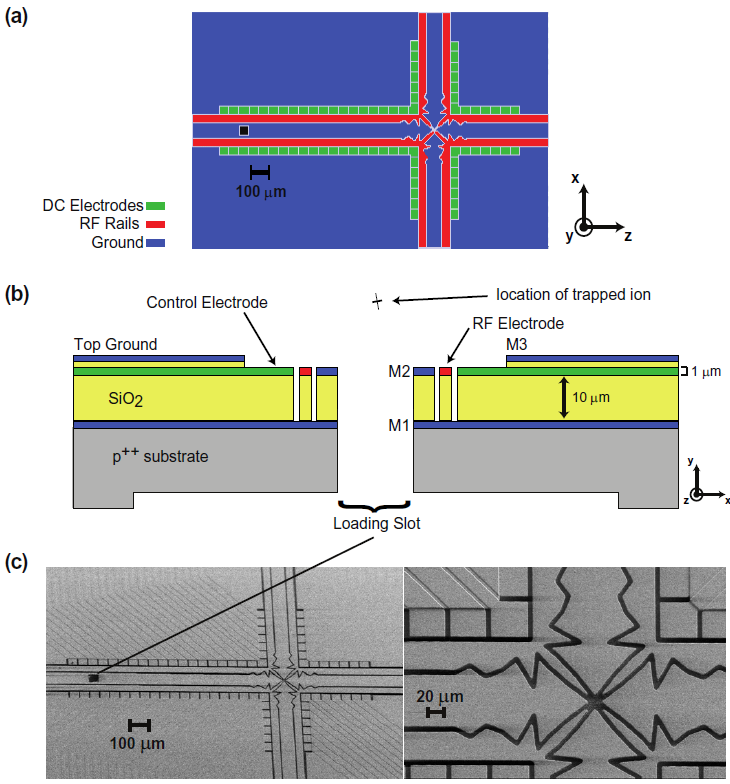}
\caption{The microfabricated X-junction surface trap used at the Georgia Tech Research Institute \cite{Wright13}. (a) shows a top-down view of the junction design with rf electrodes (red), control electrodes (green) and grounded electrodes (blue). (b) shows a cross section of the trap with aluminum layers marked M1, M2 and M3. (c) shows a scanning electron microscope image of the trap along with a detailed close-up of the junction centre.}
\label{georgia_x}
\end{figure}

Scalable ion trap architectures, particularly those aimed at quantum computing, are likely to require many electrodes to control the position of individual ions and also require junctions around which the ions will need to be shuttled \cite{Kielpinski2002}. Engineering sequences of applied electrode voltages can serve to move the ion along the trap. This allows for additional dynamical control and allows for better understanding of the extent to which quantum information can be distributed within one node without corrective measures such as sympathetic cooling \cite{Shu14}. The trap geometries in Fig.~\ref{georgia_y} and Fig.~\ref{georgia_x} show a Y- and X-junction respectively. Both of these junctions have been optimized to reduce any rf barriers located around the junctions. Shuttling around corners into different regions could be extremely important in future large scale ion trap architectures in order to create storage, interaction and ion-photon read-out zones (see also Fig.~\ref{NIST_computing}).

\begin{figure}[!htp]
\centering
\includegraphics[width=0.5\columnwidth]{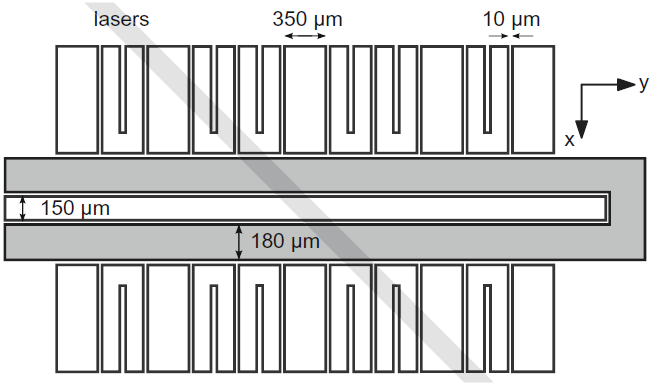}
\caption{A schematic of the surface trap structure from the University of Siegen \cite{Kunert14}. The rf electrode is shown in grey. The static voltage electrodes can be used for ion transport and the looped electrodes for currents to tailor the axial magnetic field gradient.}
\label{seigen_surface}
\end{figure}

\begin{figure}[!htp]
\centering
\includegraphics[width=0.6\columnwidth]{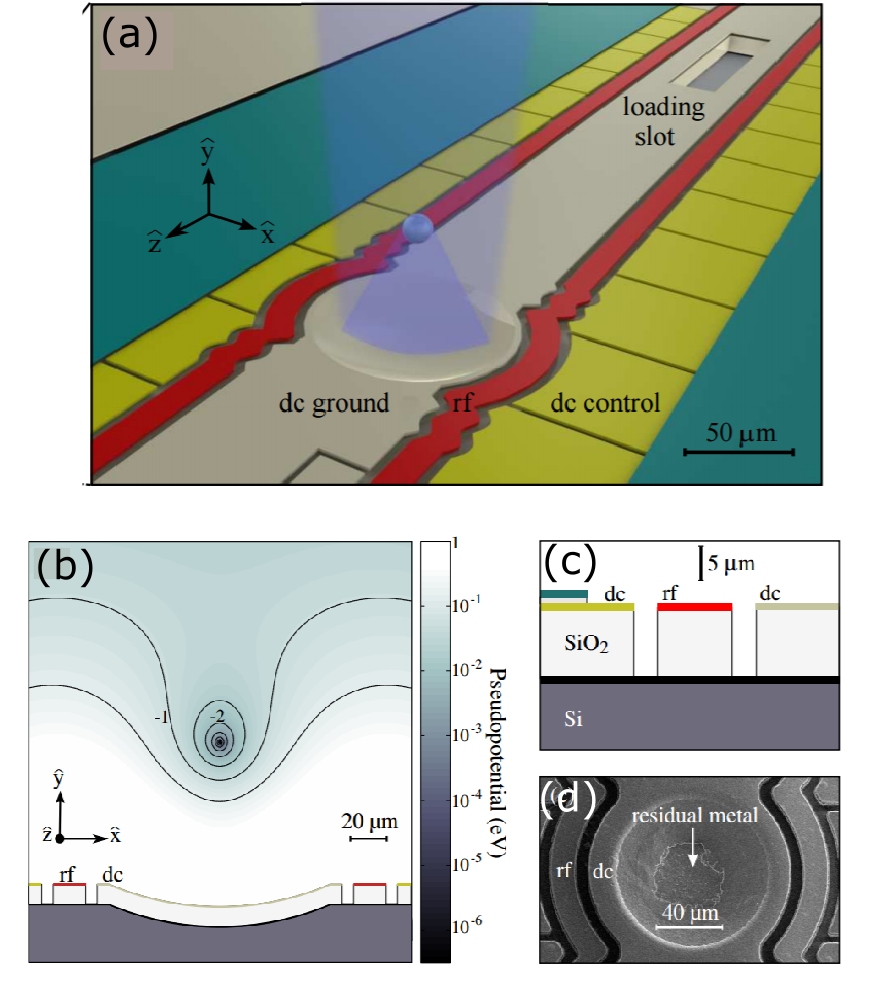}
\caption{A trap with a concave mirror embedded into the chip, used at the Georgia Institute of Technology and the Georgia Tech Research Institute \cite{Merrill11} (a) diagram of trap surface with integrated mirror and rf and dc electrodes. The rf rails are wrapped around the mirror to improve the collection efficiency (b) cross section of the trap at the center of the mirror with the pondermotive pseudopotential superimposed (b) Trap electrodes atop a 10 $\mu$m SiO$_2$ substrate which is electrically isolated from the Si substrate using a 1 $\mu$m Al ground plane. The third layer (visible on the left hand side) is an integrated capacitive filter (c) SEM image of mirror.}
\label{GTRI_mirrortrap}
\end{figure}

While most ion traps only support rf and dc potentials, there is contemporaneous work in integrating current carrying wires for controlling magnetically sensitive transitions within the ion \cite{Ospelkaus2011,Shappert13}. Figure~\ref{seigen_surface} shows a trap having segmented electrodes allowing for tailored potentials for ion shuttling but also having integrated current loops for generating magnetic field gradients with a tailorable profile and strength along the axial direction. Optics have also been integrated into microfabricated traps for enhanced photon collection (akin to the bulk trap shown in Fig.~\ref{washington_tack}), as shown in Fig.~\ref{GTRI_mirrortrap}. Here, a micro-mirror with a radius of 50.5 $\mu$m and a radius of curvature of 178 $\mu$m provides a numerical-aperture of 0.63 NA \cite{Merrill11}. With the design of the rf rails as shown in Fig.~\ref{GTRI_mirrortrap}(a)\cite{Wesenberg08}, ion transport across the mirror has been shown, demonstrating compatibility with schemes involving ion shuttling. Additionally, routing light to the ion is important for scalability and individual addressing, and one approach is to use microfabricated optical waveguides \cite{Mehta2016} which are particularly suited for large-scale quantum information processing and also for quantum networking. Integrated optics, with a microfabircated Fresnel \cite{Streed11}, is one approach that could offer a means for imaging and/or high numerical aperture photon collection. 

\begin{figure}[!htp]
\centering
\includegraphics[width=0.4\columnwidth]{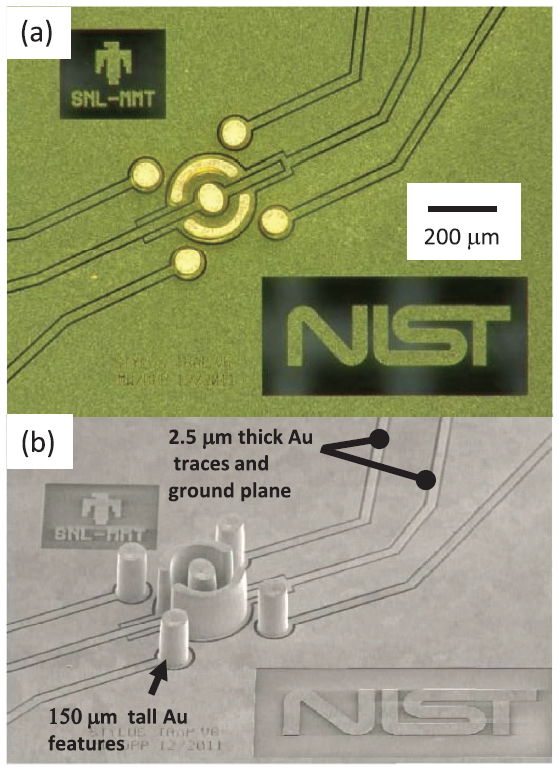}
\caption{The NIST stylus trap shown in an optical, (a), and scanning microscope, (b), images \cite{Arrington13}. The 3D nature and overall design of this trap makes it well-suited to study electric-field noise on the ion due to nearby surfaces.}
\label{nist_stylus}
\end{figure}

As microfabricated traps continue to miniaturize ion trap technology the effects of anomalous heating could, potentially, become more important. It is with this in mind that the stylus trap geometry shown in Fig.~\ref{nist_stylus} has been constructed. This trap and surrounding apparatus enables the user to place different types of surfaces and materials into close proximity of the ion to measure the effect of the surface on the heating rate of the trapped ion, as discussed in Section \ref{sec:heating}. Work of this kind betters the understanding of mechanisms of anomalous heating and allows ion traps to be made smaller and more scalable than presently available.

\begin{figure}[!htp]
\centering
\includegraphics[width=0.8\columnwidth]{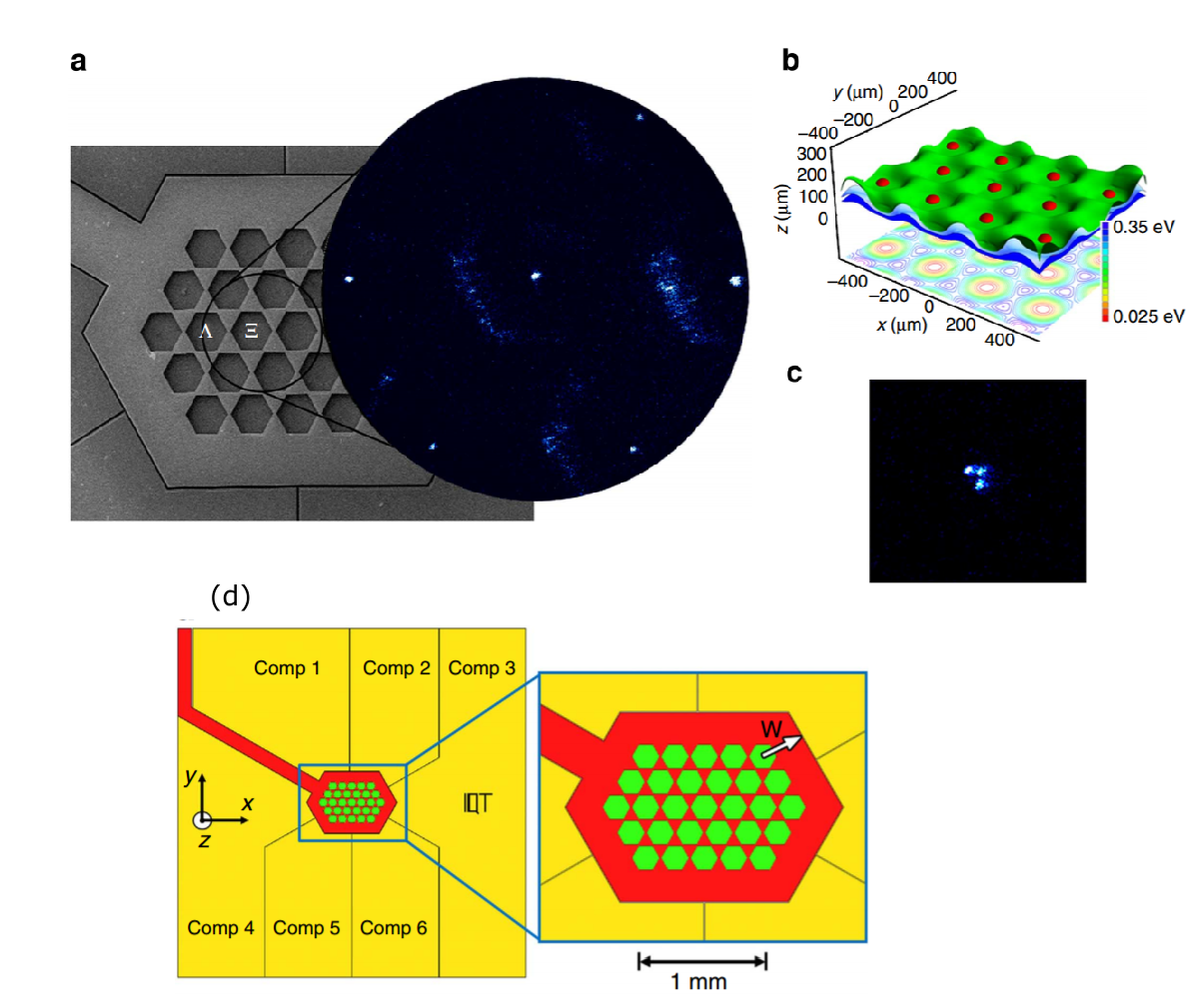}
\caption{A microfabricated 2-dimensional trap used at the University of Sussex \cite{Sterling14} (a) an SEM image of the trap and a superposed CCD image of six ions trapped in the lattice (b) three-dimensional contour plot of central 11 trap sites with an rf voltage amplitude of 445 V, equipotential surfaces correspond to: red = 0.04 eV, green = 0.4 eV, light blue = 0.5 eV and blue = 0.6 eV (c) a CCD image of a single site containing three ions (d) electrode design with 29 trapping sites, where red identifies rf electrodes, green identifies dc electrodes and yellow identifies compensation electrodes.}
\label{sussex}
\end{figure}

Two-dimensional traps have also been developed allowing the trapping of ion arrays in a lattice \cite{Sterling14,Kumph16}, with one example shown in Fig.~\ref{sussex}. With careful design it is possible for high values of the rf amplitude (455 V) to be applied \cite{Sterling14}, resulting in a relatively strong trapping field for microfabricated traps given the quadratic dependence of the pseudo-potential on the applied rf voltage (Eqn.~\ref{paul_pot}). The array contains 29 traps in a triangular lattice configuration, where each ion has six nearest neighbors (discounting the traps on the edge) and separation of 270.5 $\mu$m. In this trap, shuttling between sites has been demonstrated, albeit with global control electrodes (the compensation electrodes shown in Fig.~\ref{sussex} (4)), rather than site-specific electrodes. Such traps have a trapping lattice similar in structure to those used to confine individual neutral atoms, although in the neutral atom case, optical lattices generate the potential wells \cite{Bloch05}. This 2D ion version offers a promising path for varied applications from quantum simulation to quantum computing, although it would be challenging to use this for quantum networking because of the limited optical access.  

\begin{figure}[!htp]
\centering
\includegraphics[width=0.4\columnwidth]{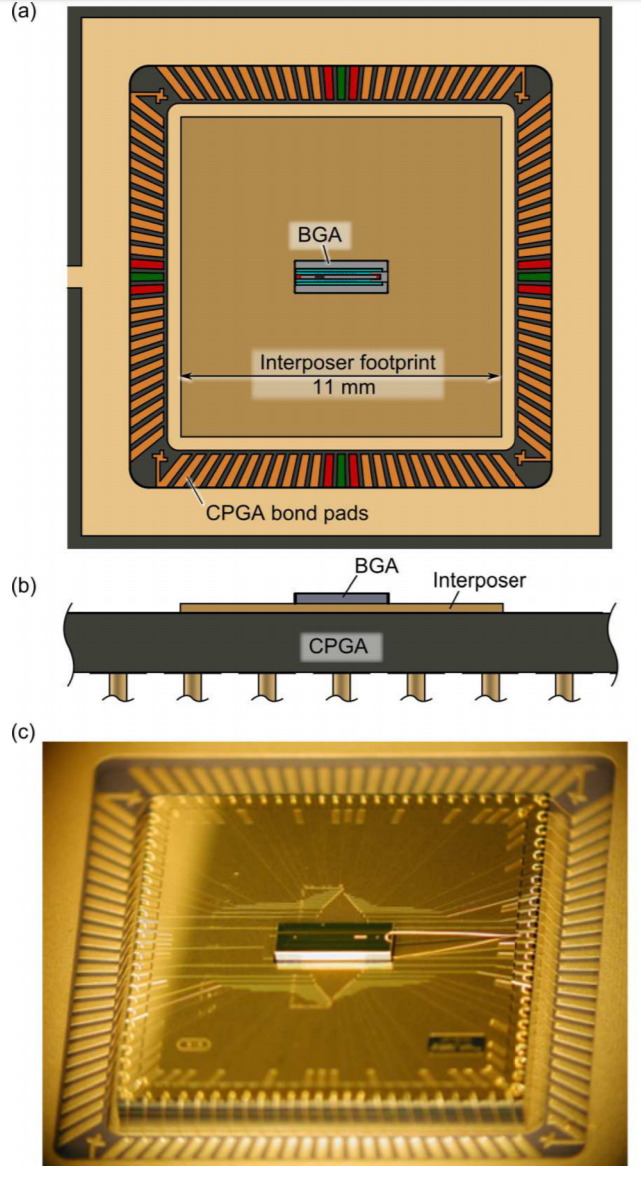}
\caption{Microfabricated surface trap from the Georgia Tech Research Institute \cite{Guise15}. The ball-grid array (BGA) allows for improved scalability as planar capacitors are fabricated into the trap die and wirebonds are moved to the interposer. The interposer sits atop a commercial ceramic pin-grid array (CPGA) carrier. The interposer has an 11 mm footprint and the trap die has 48 trench, sub-surface, capacitors and 48 DC electrodes. This geometry allows for less obstruction around the trapping region, allowing for tighter focusing for qubit operations.}
\label{GTRI_trap}
\end{figure}

Interest in scalability and robust performance has lead to the development of foundries for ion trap fabrication. Such efforts are dedicated to development and refinement of versatile platforms, designed via interactions between the community and the foundry. The extensive expertise at ion trap foundries in material science have lead to significant engineering advances. A major advantage of microfabricated traps is that they allow for improved scalability as filter capacitors are fabricated on the chip itself rather than placed in-plane. Additionally, delicate wirebond connections are re-situated on the interposer rather than extending directly from the trap, such as in \cite{Guise15} and used in the ball-grid array (BGA) trap shown in  Fig.~\ref{GTRI_trap}. Such advances are significant for scalability as some modern microfabricated traps have nearly 100 DC control electrodes \cite{Wright13} and the filter capacitors and bond pads can consume a large amount of the chip area. Future work in this area includes integration with in-vacuum electronics such as a DAC for electrode voltage control \cite{Guise14}.

\begin{table}[!ht]
\centering
		\begin{tabular}{|c|c|c|c|c|c|c|}
		\hline
		Trap & RF Drive & RF & Secular & Trap & Ion-electrode & Ion \\
		     & Frequency & Voltage & Frequency & Depth (eV) & distance &  Species\\
		          & $\Omega_T/2\pi$  & $V_0$ & $\omega/2\pi$ & $T_D$ & $r$ &  \\
        \hline
        NIST & 62.2 MHz & 85 V &8 MHz (axial) & 0.1  &  62 $\mu$m & Mg$^+$ \\
        stylus \cite{Arrington13} &&& 4 MHz (radial)&&&\\
        \hline
        Sandia & 45 MHz & $\approx$ 100 V & 1 - 1.8 MHz (axial) & various &  70 $\mu$m & Ca$^+$ \\
        Y \cite{Shu14} &&& 4 - 6 MHz (radial) &&&\\
        \hline
        GTRI & 58.55 MHz & 129 V & 0.5 - 1 MHz (axial) & 0.029  &  40 - 65 $\mu$m & Ca$^+$ \\
        X \cite{Wright13} &&& 1 - 2.5 MHz (radial) &&&\\
        \hline
        Siegen & 14.7 MHz & 125 V & 180 - 250 kHz (axial) & 0.073  &  160 $\mu$m & Yb$^+$ \\
        surface \cite{Kunert14} &&& 1.0 - 1.8 MHz (radial) &&&\\
				\hline
				GTRI & 62.3 MHz  & 200 V & 1.0 MHz (axial) & 0.01-0.10 & 178$\mu$m & Ca$+$\\
				mirror \cite{Merrill11} &&&   2.9/2.2 MHz (radial)  &&&\\
				\hline
				Sussex \cite{Sterling14} & 32.2 MHz & 455 V & ($\omega_x,\omega_y,\omega_z$)/2$\pi$ = (1.58, 1.47, 3.30) MHz & 0.42 eV & 156 $\mu$m & Yb$^+$ \\
				\hline
				GTRI & 55.14 MHz & 95 V & 1 MHz (axial) &  0.060- 0.100 \cite{Brown} & 60 $\mu$m & Yb$^+$, Ca$^+$\\
				BGA \cite{Guise15} &&& 3.7 MHz, 4.2 MHz (radial) &&& \\
				\hline
				Sandia  & 45 MHz & 250 V & 500 kHz (axial)   & 0.06  & 70 $\mu$m (slot) & Yb$^+$ \\
				HOA-2 \cite{Maunz16} 	&&& 2 MHz (radial) && 85 $\mu$m (junction) & \\
        \hline
        \end{tabular}\label{tbl:fab_asym_traps}
\caption{Summary of typical operating specifications of the microfabricated asymmetric Paul traps described in the text.}
\end{table}

\begin{figure}[!htp]
\centering
\includegraphics[width=0.4\columnwidth]{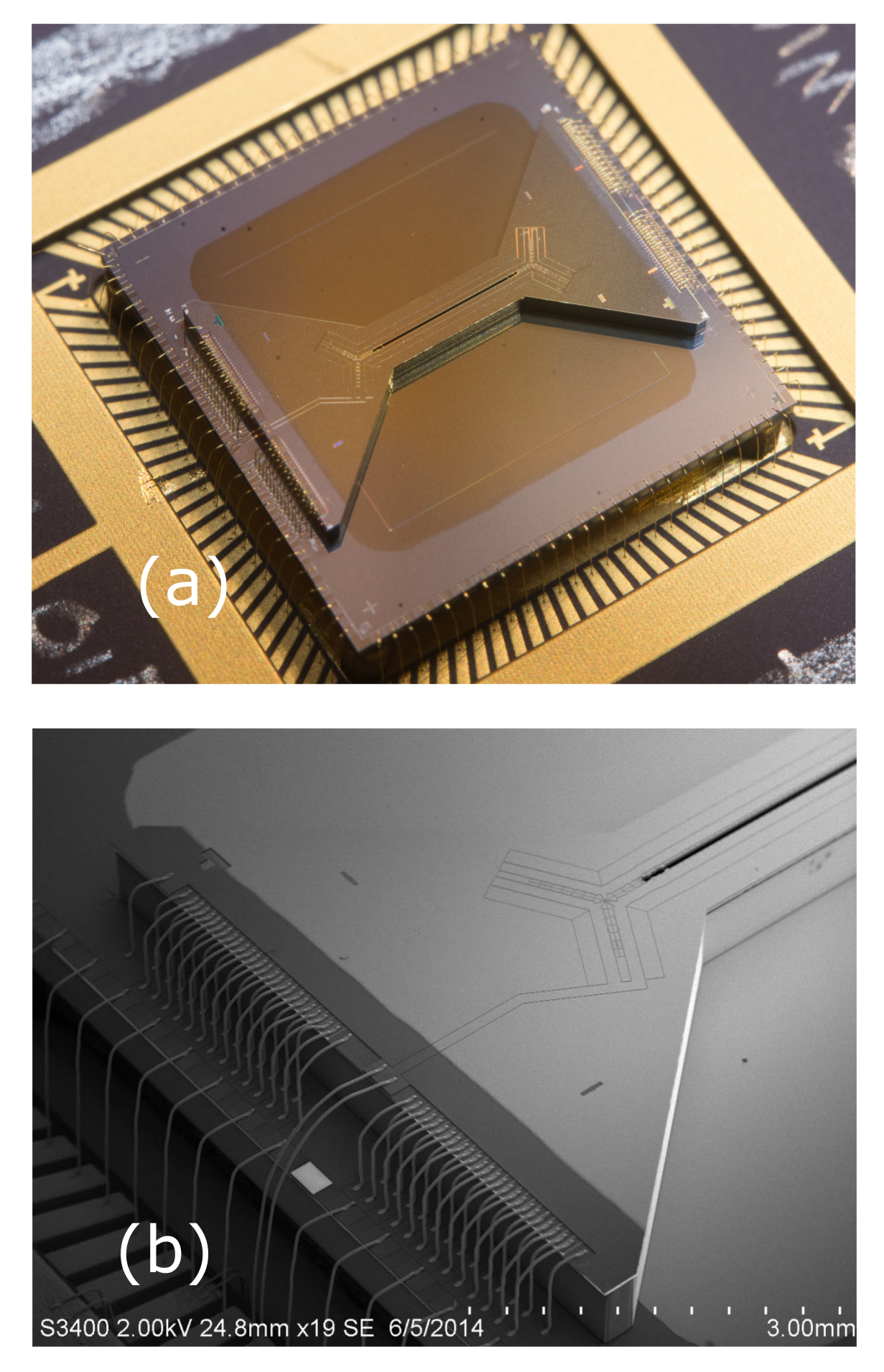}
\caption{Microfabricated High Optical Access (HOA-2) surface trap from Sandia National Labs \cite{Maunz16} where the Y-junction and slot region are visible (a) photo of fully assembled trap seated on chip carrier (b) scanning electron micrograph image of the trap showing wirebonds connecting 38 dc control signals and two wirebonds which connect directly to the trap for applying the rf trapping voltage.  This trap has zone-specific features combining Y-junctions (see Fig.~\ref{georgia_y}) and a slot (60 $\mu$m opening and 1.2 mm in length) for high optical access and ion reconfigurability.}
\label{Sandia_trap}
\end{figure}

A multi-zone, single slot approach, as shown in Fig.~\ref{Sandia_trap}, maximizes optical access with zone-specific shuttling and reconfigurable user-defined potentials. In this trap, there are dedicated regions for loading, ion re-ordering (akin to the Y-junction shown in Fig.~\ref{georgia_y}\cite{Shu14}), surface-to-slot region, shuttling region and gate/control (slot) region. The slot provides high optical access (HOA) for transverse addressing of the ions and the overall design allows for tight focusing across the surface. Similar provisions as in Fig.~\ref{GTRI_trap} are taken in to account for capacitor integration and wirebond placement. The slot width is 60 $\mu$m and allowing a beam with a waist of several microns to be focused on an ion. Remarkable results have been achieved with the latest microfabricated traps, with trapping times of days (reducing the amount of time spent reloading the trap) and continuous measurement times of hours (allowing for better averaging statistics, systematic studies of performance) \cite{Maunz16-2}. The slot could potentially allow for placement of both an imaging optic needed for diagnosis/trapping and a high-NA optic for optimal single photon collection. Table \ref{tbl:fab_asym_traps} shows the salient characteristics of the traps discussed in this section.

%-------------------------------------------------------------------------

%-------------------------------------------------------------------------
\section{Outlook}
\begin{figure}[!htp]
\centering
\includegraphics[width=0.6\columnwidth]{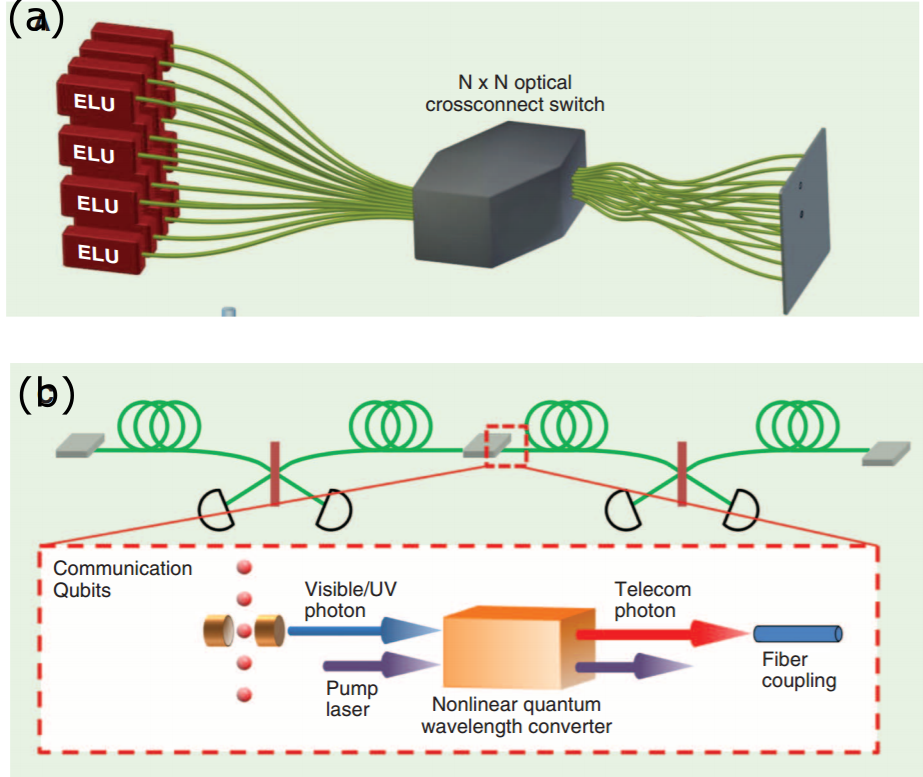}
\caption{Platform for quantum networking with trapped ions \cite{Monroe1164}(a) where elementary logic units (ELU) are connected via switch and then detected by an array of detectors and (b) long-distance transmission of photons from trapped ions via quantum frequency conversion and entanglement generation via a Bell-state measurement \cite{Matsukevich08}.}
\label{Monroe_networking}
\end{figure}

\begin{figure}[!htp]
\centering
\includegraphics[width=0.5\columnwidth]{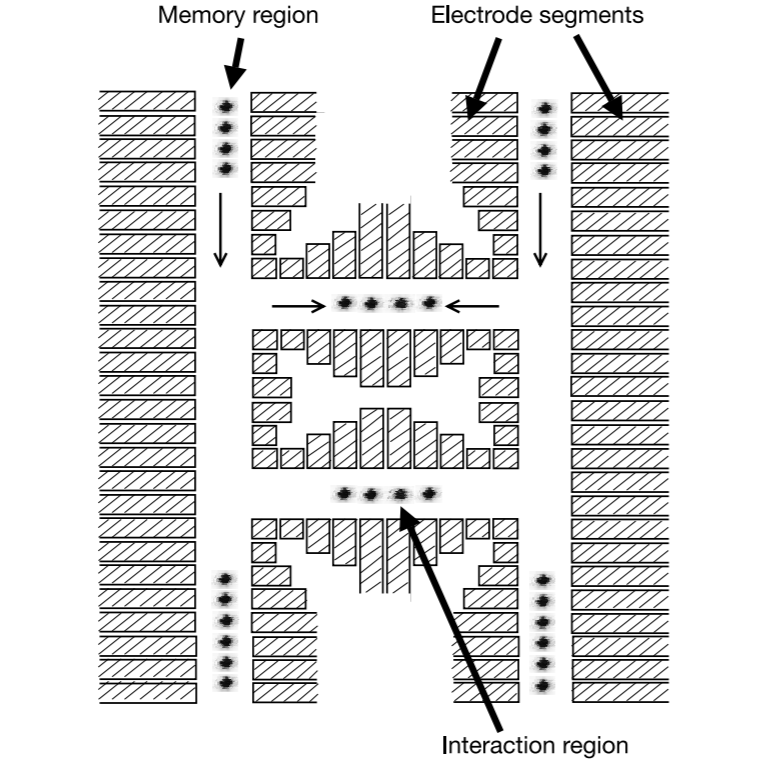}
\caption{Platform for quantum computing using trapped ions and shuttling between zones, where memory regions store the information and interaction regions interface the memories. Sequences of voltages on the electrodes serves to shuttle the ions between zones \cite{Kielpinski2002}.}
\label{NIST_computing}
\end{figure}

In recent years, efforts have been made to replicate the results observed in bulk traps on the platform of microfabricated traps. Although bulk traps still serve the community well \cite{Debnath2016,Martinez2016,Gaebler16}, microfabrication has opened new possibilities in scalability and control. The field has evolved to include hybrid approaches (such as cryogenic ion traps \cite{Brandl16,Antohi09,Vittorini13}, SNSPDs \cite{Slichter16} and microfabricated quantum interfaces) to meet challenges in studies of fundamental quantum dynamics, precision time-keeping, quantum simulation, quantum networking (see Fig.~\ref{Monroe_networking}) and quantum computation (see Fig.~\ref{NIST_computing}).  Significant engineering challenges have been overcome and many-hour lifetimes \cite{Maunz16-2}, quantum gates \cite{Herold16} and integrated optics \cite{Merrill11,Ghadimi16} have been demonstrated in microfabricated traps.  

Along-side the progress in fabrication has been a parallel effort in software development for controlling electrodes, sequencing the experiment and data processing (see this Issue's article by E. Mount, et. al \cite{Mount2015}). While individual groups have well-articulated sequencing programs, there are efforts underway to develop a comprehensive, multi-function package that incorporates often-used functionality (see this Issue's article by K. E. Stevens, et. al). Such software can readily integrate complex conditional logic operations, utilizing versatile technology based on field-programmable gate arrays (FPGA). There are now publicly available codes aimed for trapped ion experiments \cite{nist_fpga,sandia_fpga}. 

The work reported here continues to be built on and promising future directions include, fundamental theoretical studies in multi-ion simulation and quantum dynamics (see this Issue's article by G-D Lin and L. M. Duan \cite{Lin2015}), hybridization (see this Issue's article by J. Wright, et.~al. \cite{Wright2016}), further improvements in time keeping \cite{Chou1630}, integrated optics \cite{Ghadimi16} (see this Issue's article by D. Kielpinski, et.~al. \cite{Kielpinski2015}), scalable ion trap development (see this Issue's article by A. M. Eltony, et.~al.\cite{Eltony2016}) and more. Although there are significant challenges ahead, the results summarized here demonstrate that many hurdles have already been overcome and substantial progress has already been made.

\noindent
Acknowledgement: \\
We thank Ken Wright and Paul Hess for a thorough reading of the manuscript and Vikas Anant for the modelling of the SNSPD. Funding provided by the Army Research Lab, Cooperative Agreement and the Center for Distributed Quantum Information. All images preprinted from publications have been licensed for use and additionally permission requested from an author to reproduce the image. The US government neither endorses nor guarantees in any way organizations, companies or products included in this article, and such mention is only given for illustrative purposes, other competing options may be equal or better than those mentioned here.

%-------------------------------------------------------------------------
\bibliography{ion_traps}{}

\begin{thebibliography}{100}

\bibitem{Paul90}
Wolfgang Paul.
\newblock Electromagnetic traps for charged and neutral particles.
\newblock {\em Rev. Mod. Phys.}, 62:531--540, Jul 1990.

\bibitem{Dehmelt67}
H.~G. Dehmelt.
\newblock Radiofrequency spectroscopy of stored ions {I}: Storage.
\newblock {\em Adv. At. Mol. Phys.}, 3:53, 1967.

\bibitem{Bollinger91}
J.J. Bollinger, D.J. Heizen, W.M. Itano, S.L. Gilbert, and D.J. Wineland.
\newblock A 303-{MHz} frequency standard based on trapped {Be$^+$} ions.
\newblock {\em Instrumentation and Measurement, IEEE Transactions on},
  40(2):126--128, April 1991.

\bibitem{Fisk97}
P.T.H. Fisk, M.J. Sellars, M.A. Lawn, and C.~Coles.
\newblock Accurate measurement of the 12.6 {GHz} "clock" transition in trapped
  {$^{171}$Yb$^+$} ions.
\newblock {\em Ultrasonics, Ferroelectrics, and Frequency Control, IEEE
  Transactions on}, 44(2):344--354, March 1997.

\bibitem{Rosenband08}
T.~Rosenband, D.~B. Hume, P.~O. Schmidt, C.~W. Chou, A.~Brusch, L.~Lorini,
  W.~H. Oskay, R.~E. Drullinger, T.~M. Fortier, J.~E. Stalnaker, S.~A. Diddams,
  W.~C. Swann, N.~R. Newbury, W.~M. Itano, D.~J. Wineland, and J.~C. Bergquist.
\newblock Frequency ratio of al+ and hg+ single-ion optical clocks; metrology
  at the 17th decimal place.
\newblock {\em Science}, 319(5871):1808--1812, 2008.

\bibitem{Huntemann16}
N.~Huntemann, C.~Sanner, B.~Lipphardt, Chr. Tamm, and E.~Peik.
\newblock Single-ion atomic clock with
  $3\ifmmode\times\else\texttimes\fi{}{10}^{-18}$ systematic uncertainty.
\newblock {\em Phys. Rev. Lett.}, 116:063001, Feb 2016.

\bibitem{Keller16}
J~Keller, T~Burgermeister, D~Kalincev, J~Kiethe, and T~E Mehlst{\"a}ubler.
\newblock Evaluation of trap-induced systematic frequency shifts for a
  multi-ion optical clock at the 10$^{-19}$ level.
\newblock {\em Journal of Physics: Conference Series}, 723(1):012027, 2016.

\bibitem{Chou1630}
C.~W. Chou, D.~B. Hume, T.~Rosenband, and D.~J. Wineland.
\newblock Optical clocks and relativity.
\newblock {\em Science}, 329(5999):1630--1633, 2010.

\bibitem{Schwartz2002}
Jae~C. Schwartz, Michael~W. Senko, and John E.~P. Syka.
\newblock A two-dimensional quadrupole ion trap mass spectrometer.
\newblock {\em Journal of the American Society for Mass Spectrometry},
  13(6):659--669, 2002.

\bibitem{Keller03}
M~Keller, B~Lange, K~Hayasaka, W~Lange, and H~Walther.
\newblock Deterministic cavity quantum electrodynamics with trapped ions.
\newblock {\em Journal of Physics B: Atomic, Molecular and Optical Physics},
  36(3):613, 2003.

\bibitem{Kreuter04}
A.~Kreuter, C.~Becher, G.~P.~T. Lancaster, A.~B. Mundt, C.~Russo, H.~H\"affner,
  C.~Roos, J.~Eschner, F.~Schmidt-Kaler, and R.~Blatt.
\newblock Spontaneous emission lifetime of a single trapped {Ca}$^{+}$ ion in a
  high finesse cavity.
\newblock {\em Phys. Rev. Lett.}, 92:203002, May 2004.

\bibitem{Barros09}
H~G Barros, A~Stute, T~E Northup, C~Russo, P~O Schmidt, and R~Blatt.
\newblock Deterministic single-photon source from a single ion.
\newblock {\em New Journal of Physics}, 11(10):103004, 2009.

\bibitem{Hiroki13}
Hiroki Takahashi, Alex Wilson, Andrew Riley-Watson, Fedja Oru\u{c}evi\'{c},
  Nicolas Seymour-Smith, Matthias Keller, and Wolfgang Lange.
\newblock An integrated fiber trap for single-ion photonics.
\newblock {\em New Journal of Physics}, 15(5):053011, 2013.

\bibitem{Odom06}
B.~Odom, D.~Hanneke, B.~D'Urso, and G.~Gabrielse.
\newblock New measurement of the electron magnetic moment using a one-electron
  quantum cyclotron.
\newblock {\em Phys. Rev. Lett.}, 97:030801, Jul 2006.

\bibitem{Porras04}
D.~Porras and J.~I. Cirac.
\newblock Effective quantum spin systems with trapped ions.
\newblock {\em Phys. Rev. Lett.}, 92:207901, May 2004.

\bibitem{Porras06}
D.~Porras and J.~I. Cirac.
\newblock Quantum manipulation of trapped ions in two dimensional coulomb
  crystals.
\newblock {\em Phys. Rev. Lett.}, 96:250501, Jun 2006.

\bibitem{Islam13}
R.~Islam, C.~Senko, W.~C. Campbell, S.~Korenblit, J.~Smith, A.~Lee, E.~E.
  Edwards, C.-C.~J. Wang, J.~K. Freericks, and C.~Monroe.
\newblock Emergence and frustration of magnetism with variable-range
  interactions in a quantum simulator.
\newblock {\em Science}, 340(6132):583--587, 2013.

\bibitem{Schindler2013}
P.~Schindler, M.~Muller, D.~Nigg, J.~T. Barreiro, E.~A. Martinez, M.~Hennrich,
  T.~Monz, S.~Diehl, P.~Zoller, and R.~Blatt.
\newblock Quantum simulation of dynamical maps with trapped ions.
\newblock {\em Nat Phys}, 9(6):361--367, Jun 2013.
\newblock Article.

\bibitem{Zhang17}
J.~Zhang, P.~W. Hess, A.~Kyprianidis, P.~Becker, A.~Lee, J.~Smith, G.~Pagano,
  I.-D. Potirniche, A.~C. Potter, A.~Vishwanath, N.~Y. Yao, and C.~Monroe.
\newblock Observation of a discrete time crystal.
\newblock {\em Nature}, 543(7644):217--220, Mar 2017.
\newblock Letter.

\bibitem{Neyenhuis16}
B.~Neyenhuis, J.~Smith, A.~C. Lee, J.~Zhang, P.~Richerme, P.~W. Hess, Z.-X.
  Gong, A.~V. Gorshkov, and C.~Monroe.
\newblock Observation of prethermalization in long-range interacting spin
  chains.
\newblock {\em arXiv:1608.00681}, 2016.

\bibitem{CiracZoller95}
J.~I. Cirac and P.~Zoller.
\newblock Quantum computations with cold trapped ions.
\newblock {\em Phys. Rev. Lett.}, 74:4091--4094, May 1995.

\bibitem{Milburn00}
G.J. Milburn, S.~Schneider, and D.F.V. James.
\newblock Ion trap quantum computing with warm ions.
\newblock {\em Fortschritte der Physik}, 48(9-11):801--810, 2000.

\bibitem{MolmerSorensen00}
Anders S\o{}rensen and Klaus M\o{}lmer.
\newblock Entanglement and quantum computation with ions in thermal motion.
\newblock {\em Phys. Rev. A}, 62:022311, Jul 2000.

\bibitem{Duan04}
L.-M. Duan.
\newblock Scaling ion trap quantum computation through fast quantum gates.
\newblock {\em Phys. Rev. Lett.}, 93:100502, Sep 2004.

\bibitem{Wineland98}
D.J. Wineland, C.~Monroe, W.~M. Itano, D.~Leibfried, B.~E. King, and D.~M.
  Meekhof.
\newblock Experimental issues in coherent quantum-state manipulation of trapped
  atomic ions.
\newblock {\em J. Res. Nat. Inst. Stand. Tech.}, 103:259--238, 1998.

\bibitem{Debnath2016}
S.~Debnath, N.~M. Linke, C.~Figgatt, K.~A. Landsman, K.~Wright, and C.~Monroe.
\newblock Demonstration of a small programmable quantum computer with atomic
  qubits.
\newblock {\em Nature}, 536(7614):63--66, Aug 2016.
\newblock Letter.

\bibitem{Monroe14}
C.~Monroe, R.~Raussendorf, A.~Ruthven, K.~R. Brown, P.~Maunz, L.-M. Duan, and
  J.~Kim.
\newblock Large-scale modular quantum-computer architecture with atomic memory
  and photonic interconnects.
\newblock {\em Phys. Rev. A}, 89:022317, Feb 2014.

\bibitem{Olmschenk07}
S.~Olmschenk, K.~C. Younge, D.~L. Moehring, D.~N. Matsukevich, P.~Maunz, and
  C.~Monroe.
\newblock Manipulation and detection of a trapped {Yb}$^{+}$ hyperfine qubit.
\newblock {\em Phys. Rev. A}, 76:052314, Nov 2007.

\bibitem{Madsen06}
M.~J. Madsen, D.~L. Moehring, P.~Maunz, R.~N. Kohn, L.-M. Duan, and C.~Monroe.
\newblock Ultrafast coherent excitation of a trapped ion qubit for fast gates
  and photon frequency qubits.
\newblock {\em Phys. Rev. Lett.}, 97:040505, Jul 2006.

\bibitem{Molmer99}
Klaus M\o{}lmer and Anders S\o{}rensen.
\newblock Multiparticle entanglement of hot trapped ions.
\newblock {\em Phys. Rev. Lett.}, 82:1835--1838, Mar 1999.

\bibitem{Blatt2008}
Rainer Blatt and David Wineland.
\newblock Entangled states of trapped atomic ions.
\newblock {\em Nature}, 453(7198):1008--1015, Jun 2008.

\bibitem{Eschner03}
J\"{u}rgen Eschner, Giovanna Morigi, Ferdinand Schmidt-Kaler, and Rainer Blatt.
\newblock Laser cooling of trapped ions.
\newblock {\em J. Opt. Soc. Am. B}, 20(5):1003--1015, May 2003.

\bibitem{Mintert01}
Florian Mintert and Christof Wunderlich.
\newblock Ion-trap quantum logic using long-wavelength radiation.
\newblock {\em Phys. Rev. Lett.}, 87:257904, Nov 2001.

\bibitem{Lake15}
K.~Lake, S.~Weidt, J.~Randall, E.~D. Standing, S.~C. Webster, and W.~K.
  Hensinger.
\newblock Generation of spin-motion entanglement in a trapped ion using
  long-wavelength radiation.
\newblock {\em Phys. Rev. A}, 91:012319, Jan 2015.

\bibitem{Hasegawa05}
T.~Hasegawa and J.~J. Bollinger.
\newblock Rotating radio frequency traps.
\newblock {\em Phys. Rev. A}, 72:043403, 2005.

\bibitem{Duan03}
L.-M. Duan and H.~J. Kimble.
\newblock Efficient engineering of multiatom entanglement through single-photon
  detections.
\newblock {\em Phys. Rev. Lett.}, 90:253601, Jun 2003.

\bibitem{Briegel98}
H.-J. Briegel, W.~D\"ur, J.~I. Cirac, and P.~Zoller.
\newblock Quantum repeaters: The role of imperfect local operations in quantum
  communication.
\newblock {\em Phys. Rev. Lett.}, 81:5932--5935, Dec 1998.

\bibitem{Gaebler16}
J.~P. Gaebler, T.~R. Tan, Y.~Lin, Y.~Wan, R.~Bowler, A.~C. Keith, S.~Glancy,
  K.~Coakley, E.~Knill, D.~Leibfried, and D.~J. Wineland.
\newblock High-fidelity universal gate set for $^{9}${Be}$^{+}$ ion qubits.
\newblock {\em Phys. Rev. Lett.}, 117:060505, Aug 2016.

\bibitem{Ballance16}
C.~J. Ballance, T.~P. Harty, N.~M. Linke, M.~A. Sepiol, and D.~M. Lucas.
\newblock High-fidelity quantum logic gates using trapped-ion hyperfine qubits.
\newblock {\em Phys. Rev. Lett.}, 117:060504, Aug 2016.

\bibitem{Moehring11}
D~L Moehring, C~Highstrete, D~Stick, K~M Fortier, R~Haltli, C~Tigges, and M~G
  Blain.
\newblock Design, fabrication and experimental demonstration of junction
  surface ion traps.
\newblock {\em New Journal of Physics}, 13(7):075018, 2011.

\bibitem{Kielpinski2002}
D.~Kielpinski, C.~Monroe, and D.~J. Wineland.
\newblock Architecture for a large-scale ion-trap quantum computer.
\newblock {\em Nature}, 417(6890):709--711, Jun 2002.

\bibitem{Hensinger06}
W.~K. Hensinger, S.~Olmschenk, D.~Stick, D.~Hucul, M.~Yeo, M.~Acton,
  L.~Deslauriers, C.~Monroe, and J.~Rabchuk.
\newblock T-junction ion trap array for two-dimensional ion shuttling, storage,
  and manipulation.
\newblock {\em Applied Physics Letters}, 88(3):034101, 2006.

\bibitem{Schug13}
M.~Schug, J.~Huwer, C.~Kurz, P.~M\"uller, and J.~Eschner.
\newblock Heralded photonic interaction between distant single ions.
\newblock {\em Phys. Rev. Lett.}, 110:213603, May 2013.

\bibitem{Kurz2014}
Christoph Kurz, Michael Schug, Pascal Eich, Jan Huwer, Philipp M{\"u}ller, and
  J{\"u}rgen Eschner.
\newblock Experimental protocol for high-fidelity heralded photon-to-atom
  quantum state transfer.
\newblock {\em Nat. Comm.}, 5:5527 EP --, Nov 2014.
\newblock Article.

\bibitem{Kimble2008}
H.~J. Kimble.
\newblock The quantum internet.
\newblock {\em Nature}, 453(7198):1023--1030, Jun 2008.

\bibitem{Wootters82}
W.~H. Wootters and W.~H. Zurek.
\newblock A single quantum cannot be cloned.
\newblock {\em Nature}, 299:802, 1982.

\bibitem{Cabrillo99}
C.~Cabrillo, J.~I. Cirac, P.~Garc\'{\i}a-Fern\'andez, and P.~Zoller.
\newblock Creation of entangled states of distant atoms by interference.
\newblock {\em Phys. Rev. A}, 59:1025--1033, Feb 1999.

\bibitem{Hucul15}
D.~Hucul, I.~V. Inlek, C.~Crocker, S.~Debnath, S.~M. Clark, and C.~Monroe.
\newblock Modular entanglement of atomic qubits using photons and phonons.
\newblock {\em Nature Physics}, 11:37--42, 2015.

\bibitem{Streed09}
Erik~W. Streed, Benjiamin~G. Norton, J.J. Chapman, and David Kielpinski.
\newblock Scalable efficient ion-photon coupling with phase fresnel lenses for
  large-scale quantum computing.
\newblock {\em Quant. Inform. Comp.}, 9:0203, 2009.

\bibitem{Siverns16}
J.~D. Siverns, Li~X., and Q.~Quraishi.
\newblock Ion-photon entanglement and quantum frequency conversion with trapped
  {Ba}$^+$ ions.
\newblock {\em Applied Physics Letters}, 56:B222, 2017.

\bibitem{Hughes11}
Marcus~D. Hughes, Bjoern Lekitsch, Jiddu~A. Broersma, and Winfried~K.
  Hensinger.
\newblock Microfabricated ion traps.
\newblock {\em Contempory Physics}, 52:505--529, 2011.

\bibitem{Turchette00}
Q.~A. Turchette, Kielpinski, B.~E. King, D.~Leibfried, D.~M. Meekhof, C.~J.
  Myatt, M.~A. Rowe, C.~A. Sackett, C.~S. Wood, W.~M. Itano, C.~Monroe, and
  D.~J. Wineland.
\newblock Heating of trapped ions from the quantum ground state.
\newblock {\em Phys. Rev. A}, 61:063418, May 2000.

\bibitem{Deslauriers06}
L.~Deslauriers, S.~Olmschenk, D.~Stick, W.~K. Hensinger, J.~Sterk, and
  C.~Monroe.
\newblock Scaling and suppression of anomalous heating in ion traps.
\newblock {\em Phys. Rev. Lett.}, 97:103007, Sep 2006.

\bibitem{hite2013}
D.A. Hite, Y.~Colombe, A.C. Wilson, D.T.C. Allcock, D.~Leibfried, D.J.
  Wineland, and D.P. Pappas.
\newblock Surface science for improved ion traps.
\newblock {\em MRS Bulletin}, 38(10):826--833, 10 2013.

\bibitem{Sage15}
Robert McConnell, Colin Bruzewicz, John Chiaverini, and Jeremy Sage.
\newblock Reduction of trapped-ion anomalous heating by \textit{in situ}
  surface plasma cleaning.
\newblock {\em Phys. Rev. A}, 92:020302, Aug 2015.

\bibitem{Allcock2012}
D.~T.~C. Allcock, T.~P. Harty, H.~A. Janacek, N.~M. Linke, C.~J. Ballance,
  A.~M. Steane, D.~M. Lucas, R.~L. Jarecki, S.~D. Habermehl, M.~G. Blain,
  D.~Stick, and D.~L. Moehring.
\newblock Heating rate and electrode charging measurements in a scalable,
  microfabricated, surface-electrode ion trap.
\newblock {\em Applied Physics B}, 107(4):913--919, 2012.

\bibitem{Olmschenk486}
S.~Olmschenk, D.~N. Matsukevich, P.~Maunz, D.~Hayes, L.-M. Duan, and C.~Monroe.
\newblock Quantum teleportation between distant matter qubits.
\newblock {\em Science}, 323(5913):486--489, 2009.

\bibitem{excelitas}
Excelitas Technologies.
\newblock {SPCM-NIR} {Rev} 2015-03.
\newblock \url{http://www.excelitas.com/downloads/SPCM-NIR_Product_Brief.pdf},
  2015.

\bibitem{Gaudio16}
Rosalinda Gaudio, Jelmer~J. Renema, Zili Zhou, Varun~B. Verma, Adriana~E. Lita,
  Jeffrey Shainline, Martin~J. Stevens, Richard~P. Mirin, Sae~Woo Nam,
  Martin~P. van Exter, Michiel J.~A. de~Dood, and Andrea Fiore.
\newblock Experimental investigation of the detection mechanism in wsi nanowire
  superconducting single photon detectors.
\newblock {\em Applied Physics Letters}, 109(3), 2016.

\bibitem{Lamas13}
Antia Lamas-Linares, Brice Calkins, Nathan~A. Tomlin, Thomas Gerrits,
  Adriana~E. Lita, J\"orn Beyer, Richard~P. Mirin, and Sae Woo~Nam.
\newblock Nanosecond-scale timing jitter for single photon detection in
  transition edge sensors.
\newblock {\em Applied Physics Letters}, 102(23), 2013.

\bibitem{Marsili13}
F.~Marsili, V.~B. Verma, J.~A. Stern, S.~Harrington, A.~E. Lita, T.~Gerrits,
  I.~Vayshenker, B.~Baek, M.~D. Shaw, R.~P. Mirin, and S.~W. Nam.
\newblock Detecting single infrared photons with 93% system efficiency.
\newblock {\em Nature Photonics}, 7(3):210--214, 03 2013.

\bibitem{Rath2015}
Patrik Rath, Oliver Kahl, Simone Ferrari, Fabian Sproll, Georgia
  Lewes-Malandrakis, Dietmar Brink, Konstantin Ilin, Michael Siegel, Christoph
  Nebel, and Wolfram Pernice.
\newblock Superconducting single-photon detectors integrated with diamond
  nanophotonic circuits.
\newblock {\em Light Sci Appl}, 4:e338, Oct 2015.
\newblock Original Article.

\bibitem{Hadfield2009}
Robert~H. Hadfield.
\newblock Single-photon detectors for optical quantum information applications.
\newblock {\em Nat Photon}, 3(12):696--705, Dec 2009.

\bibitem{Yamashita11}
T.~Yamashita, S.~Miki, K.~Makise, W.~Qiu, H.~Terai, M.~Fujiwara, M.~Sasaki, and
  Z.~Wang.
\newblock Origin of intrinsic dark count in superconducting nanowire
  single-photon detectors.
\newblock {\em Applied Physics Letters}, 99(16), 2011.

\bibitem{Anant16}
Vikas Anant.
\newblock Photon spot inc.
\newblock http://www.photonspot.com/, 2016.

\bibitem{Slichter16}
D.~H. Slichter, V.~B. Verma, D.~Liebfried, R.~P. Mirin, S.~W. Nam, and D.~J.
  Wineland.
\newblock {UV}-sensitive superconducting nanowire single photon detectors for
  integration in an ion trap.
\newblock {\em arXiv:1611.09949}, 2016.

\bibitem{Kumar90}
Prem Kumar.
\newblock Quantum frequency conversion.
\newblock {\em Opt. Lett.}, 15(24):1476--1478, Dec 1990.

\bibitem{Lenhard:17}
Andreas Lenhard, Jos\'{e} Brito, Matthias Bock, Christoph Becher, and
  J\"{u}rgen Eschner.
\newblock Coherence and entanglement preservation of frequency-converted
  heralded single photons.
\newblock {\em Opt. Express}, 25(10):11187--11199, May 2017.

\bibitem{Feyer16}
Vahid Esfandyarpour, Carsten Langrock, and M.~M. Fejer1.
\newblock Cascaded downconversion interface to the telecom band for
  single-photon-level signals at 650 nm.
\newblock Conference on Lasers and Electro-optics, 2016.

\bibitem{Li16}
Li~X., J.~D. Siverns, and Q.~Quraishi.
\newblock Hybrid ion-neutral atom network using quantum frequency conversion.
\newblock {\em To be published}, 2017.

\bibitem{Zaske12}
Sebastian Zaske, Andreas Lenhard, Christian~A. Ke\ss{}ler, Jan Kettler,
  Christian Hepp, Carsten Arend, Roland Albrecht, Wolfgang-Michael Schulz,
  Michael Jetter, Peter Michler, and Christoph Becher.
\newblock Visible-to-telecom quantum frequency conversion of light from a
  single quantum emitter.
\newblock {\em Phys. Rev. Lett.}, 109:147404, Oct 2012.

\bibitem{Madsen04}
M.J. Madsen, W.K. Hensinger, D.~Stick, J.A. Rabchuk, and C.~Monroe.
\newblock Planar ion trap geometry for microfabrication.
\newblock {\em Applied Physics B}, 78(5):639--651, 2004.

\bibitem{Ghosh}
P~K Ghosh.
\newblock {\em Ion Traps}.
\newblock Oxford University Press, 1995.

\bibitem{Berkeland98}
D.~J. Berkeland, J.~D. Miller, J.~C. Bergquist, W.~M. Itano, and D.~J.
  Wineland.
\newblock Minimization of ion micromotion in a paul trap.
\newblock {\em Journal of Applied Physics}, 83(10):5025--5033, 1998.

\bibitem{Drewsen98}
M.~Drewsen, C.~Brodersen, L.~Hornek\ae{}r, J.~S. Hangst, and J.~P. Schifffer.
\newblock Large ion crystals in a linear paul trap.
\newblock {\em Phys. Rev. Lett.}, 81:2878--2881, Oct 1998.

\bibitem{Vittorini14}
G.~Vittorini, D.~Hucul, I.~V. Inlek, C.~Crocker, and C.~Monroe.
\newblock Entanglement of distinguishable quantum memories.
\newblock {\em Phys. Rev. A}, 90:040302, Oct 2014.

\bibitem{Clark14}
Craig~R. Clark, Chin-wen Chou, A.~R. Ellis, Jeff Hunker, Shanalyn~A. Kemme,
  Peter Maunz, Boyan Tabakov, Chris Tigges, and Daniel~L. Stick.
\newblock Characterization of fluorescence collection optics integrated with a
  microfabricated surface electrode ion trap.
\newblock {\em Phys. Rev. Applied}, 1:024004, Mar 2014.

\bibitem{Tabakov15}
Boyan Tabakov, Francisco Benito, Matthew Blain, Craig~R. Clark, Susan Clark,
  Raymond~A. Haltli, Peter Maunz, Jonathan~D. Sterk, Chris Tigges, and Daniel
  Stick.
\newblock Assembling a ring-shaped crystal in a microfabricated surface ion
  trap.
\newblock {\em Phys. Rev. Applied}, 4:031001, Sep 2015.

\bibitem{Doret12}
S~Charles Doret, Jason~M Amini, Kenneth Wright, Curtis Volin, Tyler Killian,
  Arkadas Ozakin, Douglas Denison, Harley Hayden, C-S Pai, Richart~E Slusher,
  and Alexa~W Harter.
\newblock Controlling trapping potentials and stray electric fields in a
  microfabricated ion trap through design and compensation.
\newblock {\em New Journal of Physics}, 14(7):073012, 2012.

\bibitem{Shu14}
G.~Shu, G.~Vittorini, A.~Buikema, C.~S. Nichols, C.~Volin, D.~Stick, and
  Kenneth~R. Brown.
\newblock Heating rates and ion-motion control in a {Y}-junction
  surface-electrode trap.
\newblock {\em Phys. Rev. A}, 89:062308, Jun 2014.

\bibitem{Herold16}
C~D Herold, S~D Fallek, J~T Merrill, A~M Meier, K~R Brown, C~E Volin, and J~M
  Amini.
\newblock Universal control of ion qubits in a scalable microfabricated planar
  trap.
\newblock {\em New Journal of Physics}, 18(2):023048, 2016.

\bibitem{Antohi09}
P.~B. Antohi, D.~Schuster, G.~M. Akselrod, J.~Labaziewicz, Y.~Ge, Z.~Lin, W.~S.
  Bakr, and I.~L. Chuang.
\newblock Cryogenic ion trapping systems with surface-electrode traps.
\newblock {\em Review of Scientific Instruments}, 80(1):013103, 2009.

\bibitem{Vittorini13}
Grahame Vittorini, Kenneth Wright, Kenneth~R. Brown, Alexa~W. Harter, and
  S.~Charles Doret.
\newblock Modular cryostat for ion trapping with surface-electrode ion traps.
\newblock {\em Review of Scientific Instruments}, 84(4):043112, 2013.

\bibitem{Labaziewicz08}
Jaroslaw Labaziewicz, Yufei Ge, Paul Antohi, David Leibrandt, Kenneth~R. Brown,
  and Isaac~L. Chuang.
\newblock Suppression of heating rates in cryogenic surface-electrode ion
  traps.
\newblock {\em Phys. Rev. Lett.}, 100:013001, Jan 2008.

\bibitem{Niedermayr14}
Michael Niedermayr, Kirill Lakhmanskiy, Muir Kumph, Stefan Partel, Jonannes
  Edlinger, Michael Brownnutt, and Rainer Blatt.
\newblock Cryogenic surface ion trap based on intrinsic silicon.
\newblock {\em New Journal of Physics}, 16(11):113068, 2014.

\bibitem{Urabe05}
Tadashi Furukawa, Jin Nishimura, Utako Tanaka, and Shinji Urabe.
\newblock Design and characteristic measurement of miniature three-segment
  linear paul trap.
\newblock {\em Japanese Journal of Applied Physics}, 44(10R):7619, 2005.

\bibitem{Bergquist86}
J.~C. Bergquist, Randall~G. Hulet, Wayne~M. Itano, and D.~J. Wineland.
\newblock Observation of quantum jumps in a single atom.
\newblock {\em Phys. Rev. Lett.}, 57:1699--1702, Oct 1986.

\bibitem{Wineland13}
David~J. Wineland.
\newblock Nobel lecture: Superposition, entanglement, and raising
  {Schr\"odinger's} cat*.
\newblock {\em Rev. Mod. Phys.}, 85:1103--1114, Jul 2013.

\bibitem{Smith2016}
J.~Smith, A.~Lee, P.~Richerme, B.~Neyenhuis, P.~W. Hess, P.~Hauke, M.~Heyl,
  D.~A. Huse, and C.~Monroe.
\newblock Many-body localization in a quantum simulator with programmable
  random disorder.
\newblock {\em Nat Phys}, 12(10):907--911, Oct 2016.
\newblock Letter.

\bibitem{Monz11}
Thomas Monz, Philipp Schindler, Julio~T. Barreiro, Michael Chwalla, Daniel
  Nigg, William~A. Coish, Maximilian Harlander, Wolfgang H\"ansel, Markus
  Hennrich, and Rainer Blatt.
\newblock 14-qubit entanglement: Creation and coherence.
\newblock {\em Phys. Rev. Lett.}, 106:130506, Mar 2011.

\bibitem{Schmidt-Kaler2003}
Ferdinand Schmidt-Kaler, Hartmut Haffner, Mark Riebe, Stephan Gulde, Gavin
  P.~T. Lancaster, Thomas Deuschle, Christoph Becher, Christian~F. Roos, Jurgen
  Eschner, and Rainer Blatt.
\newblock Realization of the {Cirac-Zoller} controlled-{NOT} quantum gate.
\newblock {\em Nature}, 422(6930):408--411, Mar 2003.

\bibitem{Monz09}
T.~Monz, K.~Kim, W.~H\"ansel, M.~Riebe, A.~S. Villar, P.~Schindler, M.~Chwalla,
  M.~Hennrich, and R.~Blatt.
\newblock Realization of the quantum {Toffoli} gate with trapped ions.
\newblock {\em Phys. Rev. Lett.}, 102:040501, Jan 2009.

\bibitem{Haffner2005}
H.~Haffner, W.~Hansel, C.~F. Roos, J.~Benhelm, D.~Chek-al kar, M.~Chwalla,
  T.~Korber, U.~D. Rapol, M.~Riebe, P.~O. Schmidt, C.~Becher, O.~Guhne, W.~Dur,
  and R.~Blatt.
\newblock Scalable multiparticle entanglement of trapped ions.
\newblock {\em Nature}, 438(7068):643--646, Dec 2005.

\bibitem{SchmidtKaler03}
F.~Schmidt-Kaler, H.~H\"affner, S.~Gulde, M.~Riebe, G.P.T. Lancaster,
  T.~Deuschle, C.~Becher, W.~H\"ansel, J.~Eschner, C.F. Roos, and R.~Blatt.
\newblock How to realize a universal quantum gate with trapped ions.
\newblock {\em Applied Physics B}, 77(8):789--796, 2003.

\bibitem{Rowe02}
M.~A. Rowe, A.~Ben-Kish, B.~Demarco, D.~Leibfried, V.~Meyer, J.~Beall,
  J.~Britton, J.~Hughes, W.~M. Itano, B.~Jelenkovi\'{c}, C.~Langer,
  T.~Rosenband, and D.~J. Wineland.
\newblock Transport of quantum states and separation of ions in a dual rf ion
  trap.
\newblock {\em Quantum Info. Comput.}, 2(4):257--271, June 2002.

\bibitem{Chiaverini2004}
J.~Chiaverini, D.~Leibfried, T.~Schaetz, M.~D. Barrett, R.~B. Blakestad,
  J.~Britton, W.~M. Itano, J.~D. Jost, E.~Knill, C.~Langer, R.~Ozeri, and D.~J.
  Wineland.
\newblock Realization of quantum error correction.
\newblock {\em Nature}, 432(7017):602--605, Dec 2004.

\bibitem{Nizamani2012}
A.~H. Nizamani and W.~K. Hensinger.
\newblock Optimum electrode configurations for fast ion separation in
  microfabricated surface ion traps.
\newblock {\em Applied Physics B}, 106(2):327--338, 2012.

\bibitem{McLoughlin11}
James~J. McLoughlin, Altaf~H. Nizamani, James~D. Siverns, Robin~C. Sterling,
  Marcus~D. Hughes, Bjoern Lekitsch, Bj\"orn Stein, Seb Weidt, and Winfried~K.
  Hensinger.
\newblock Versatile ytterbium ion trap experiment for operation of scalable
  ion-trap chips with motional heating and transition-frequency measurements.
\newblock {\em Phys. Rev. A}, 83:013406, Jan 2011.

\bibitem{Korenblit12}
S~Korenblit, D~Kafri, W~C Campbell, R~Islam, E~E Edwards, Z-X Gong, G-D Lin,
  L-M Duan, J~Kim, K~Kim, and C~Monroe.
\newblock Quantum simulation of spin models on an arbitrary lattice with
  trapped ions.
\newblock {\em New Journal of Physics}, 14(9):095024, 2012.

\bibitem{Kim2010}
K.~Kim, M.-S. Chang, S.~Korenblit, R.~Islam, E.~E. Edwards, J.~K. Freericks,
  G.-D. Lin, L.-M. Duan, and C.~Monroe.
\newblock Quantum simulation of frustrated {Ising} spins with trapped ions.
\newblock {\em Nature}, 465(7298):590--593, Jun 2010.

\bibitem{Edwards10}
E.~E. Edwards, S.~Korenblit, K.~Kim, R.~Islam, M.-S. Chang, J.~K. Freericks,
  G.-D. Lin, L.-M. Duan, and C.~Monroe.
\newblock Quantum simulation and phase diagram of the transverse-field {Ising}
  model with three atomic spins.
\newblock {\em Phys. Rev. B}, 82:060412, Aug 2010.

\bibitem{Islam2011}
R.~Islam, E.~E. Edwards, K.~Kim, S.~Korenblit, C.~Noh, H.~Carmichael, G.-D.
  Lin, L.-M. Duan, C.-C. Joseph~Wang, J.~K. Freericks, and C.~Monroe.
\newblock Onset of a quantum phase transition with a trapped ion quantum
  simulator.
\newblock {\em Nature Communications}, 2:377 EP --, Jul 2011.
\newblock Article.

\bibitem{Senko430}
C.~Senko, J.~Smith, P.~Richerme, A.~Lee, W.~C. Campbell, and C.~Monroe.
\newblock Coherent imaging spectroscopy of a quantum many-body spin system.
\newblock {\em Science}, 345(6195):430--433, 2014.

\bibitem{Pyka2014}
Karsten Pyka, Norbert Herschbach, Jonas Keller, and Tanja~E. Mehlst{\"a}ubler.
\newblock A high-precision segmented {Paul} trap with minimized micromotion for
  an optical multiple-ion clock.
\newblock {\em Applied Physics B}, 114(1):231--241, 2014.

\bibitem{Deslauriers04}
L.~Deslauriers, P.~C. Haljan, P.~J. Lee, K-A. Brickman, B.~B. Blinov, M.~J.
  Madsen, and C.~Monroe.
\newblock Zero-point cooling and low heating of trapped $^{111}${Cd}$^{+}$
  ions.
\newblock {\em Phys. Rev. A}, 70:043408, Oct 2004.

\bibitem{Seb15}
S.~Weidt, J.~Randall, S.~C. Webster, E.~D. Standing, A.~Rodriguez, A.~E. Webb,
  B.~Lekitsch, and W.~K. Hensinger.
\newblock Ground-state cooling of a trapped ion using long-wavelength
  radiation.
\newblock {\em Phys. Rev. Lett.}, 115:013002, Jun 2015.

\bibitem{Weidt16}
S.~Weidt, J.~Randall, S.~C. Webster, K.~Lake, A.~E. Webb, I.~Cohen,
  T.~Navickas, B.~Lekitsch, A.~Retzker, and W.~K. Hensinger.
\newblock Trapped-ion quantum logic with global radiation fields.
\newblock {\em Phys. Rev. Lett.}, 117:220501, Nov 2016.

\bibitem{Brandstatter13}
B.~Brandst\"atter, A.~McClung, K.~Sch\"uppert, B.~Casabone, K.~Friebe,
  A.~Stute, P.~O. Schmidt, C.~Deutsch, J.~Reichel, R.~Blatt, and T.~E. Northup.
\newblock Integrated fiber-mirror ion trap for strong ion-cavity coupling.
\newblock {\em Review of Scientific Instruments}, 84(12):123104, 2013.

\bibitem{Streed11}
Erik~W. Streed, Benjamin~G. Norton, Andreas Jechow, Till~J. Weinhold, and David
  Kielpinski.
\newblock Imaging of trapped ions with a microfabricated optic for quantum
  information processing.
\newblock {\em Phys. Rev. Lett.}, 106:010502, 2011.

\bibitem{Casabone15}
B.~Casabone, K.~Friebe, B.~Brandst\"atter, K.~Sch\"uppert, R.~Blatt, and T.~E.
  Northup.
\newblock Enhanced quantum interface with collective ion-cavity coupling.
\newblock {\em Phys. Rev. Lett.}, 114:023602, Jan 2015.

\bibitem{Vogell17}
B.~Vogell, B.~Vermersch, T.~E. Northup, B.~P. Lanyon, and C.~A. Muschik.
\newblock Deterministic quantum state transfer between remote qubits in
  cavities.
\newblock {\em arXiv:1704.06233v2}, 2017.

\bibitem{Sterk12}
J.~D. Sterk, L.~Luo, T.~A. Manning, P.~Maunz, and C.~Monroe.
\newblock Photon collection from a trapped ion-cavity system.
\newblock {\em Phys. Rev. A}, 85:062308, Jun 2012.

\bibitem{Steiner13}
Matthias Steiner, Hendrik~M. Meyer, Christian Deutsch, Jakob Reichel, and
  Michael K\"ohl.
\newblock Single ion coupled to an optical fiber cavity.
\newblock {\em Phys. Rev. Lett.}, 110:043003, Jan 2013.

\bibitem{Takahashi17}
H.~Takahashi, E.~Kassa, C.~Christoforou, and M.~Keller.
\newblock Cavity-induced anti-correlated photon emission rates of a single ion.
\newblock {\em arXiv:1705.07923}, 2017.

\bibitem{Meyer15}
H.~M. Meyer, R.~Stockill, M.~Steiner, C.~Le~Gall, C.~Matthiesen, E.~Clarke,
  A.~Ludwig, J.~Reichel, M.~Atat\"ure, and M.~K\"ohl.
\newblock Direct photonic coupling of a semiconductor quantum dot and a trapped
  ion.
\newblock {\em Phys. Rev. Lett.}, 114:123001, Mar 2015.

\bibitem{Wright2016}
John Wright, Carolyn Auchter, Chen-Kuan Chou, Richard~D. Graham, Thomas~W.
  Noel, Tomasz Sakrejda, Zichao Zhou, and Boris~B. Blinov.
\newblock Toward a scalable quantum computing architecture with mixed species
  ion chains.
\newblock {\em Quantum Information Processing}, pages 1--11, 2016.

\bibitem{Begley16}
Stephen Begley, Markus Vogt, Gurpreet~Kaur Gulati, Hiroki Takahashi, and
  Matthias Keller.
\newblock Optimized multi-ion cavity coupling.
\newblock {\em Phys. Rev. Lett.}, 116:223001, May 2016.

\bibitem{Fogarty15}
T.~Fogarty, C.~Cormick, H.~Landa, Vladimir~M. Stojanovi\ifmmode~\acute{c}\else
  \'{c}\fi{}, E.~Demler, and Giovanna Morigi.
\newblock Nanofriction in cavity quantum electrodynamics.
\newblock {\em Phys. Rev. Lett.}, 115:233602, Dec 2015.

\bibitem{Shu11}
Gang Shu, Chen-Kuan Chou, Nathan Kurz, Matthew~R. Dietrich, and Boris~B.
  Blinov.
\newblock Efficient fluorescence collection and ion imaging with the ``tack''
  ion trap.
\newblock {\em J. Opt. Soc. Am. B}, 28(12):2865--2870, Dec 2011.

\bibitem{Maiwald12}
Robert Maiwald, Andrea Golla, Martin Fischer, Marianne Bader, Simon Heugel,
  Beno\^{\i}t Chalopin, Markus Sondermann, and Gerd Leuchs.
\newblock Collecting more than half the fluorescence photons from a single ion.
\newblock {\em Phys. Rev. A}, 86:043431, Oct 2012.

\bibitem{Kumph16}
M~Kumph, P~Holz, K~Langer, M~Meraner, M~Niedermayr, M~Brownnutt, and R~Blatt.
\newblock Operation of a planar-electrode ion-trap array with adjustable rf
  electrodes.
\newblock {\em New Journal of Physics}, 18(2):023047, 2016.

\bibitem{Siverns12}
James~D Siverns, Seb Weidt, Kim Lake, Bjoern Lekitsch, Marcus~D Hughes, and
  Winfried~K Hensinger.
\newblock Optimization of two-dimensional ion trap arrays for quantum
  simulation.
\newblock {\em New Journal of Physics}, 14(8):085009, 2012.

\bibitem{Sterling14}
R.~C. Sterling, H.~Rattanasonti, S.~Weidt, K.~Lake, P.~Srinivasan, S.~C.
  Webster, M.~Kraft, and W.~K. Hensinger.
\newblock Fabrication and operation of a two-dimensional ion-trap lattice on a
  high-voltage microchip.
\newblock {\em Nature Communications}, 5:3637 EP --, Apr 2014.
\newblock Article.

\bibitem{Lindenfelser15}
F.~Lindenfelser, B.~Keitch, D.~Kienzler, D.~Bykov, P.~Uebel, M.~A. Schmidt,
  P.~St.~J. Russell, and J.~P. Home.
\newblock An ion trap built with photonic crystal fibre technology.
\newblock {\em Review of Scientific Instruments}, 86(3):033107, 2015.

\bibitem{Stick05}
D.~Stick, W.~K. Hensinger, S.~Olmschenk, M.~J. Madsen, K.~Schwab, and
  C.~Monroe.
\newblock Ion trap in a semiconductor chip.
\newblock {\em Nature Physics}, 2:36--39, 2006.

\bibitem{Wright13}
Kenneth Wright, Jason~M Amini, Daniel~L Faircloth, Curtis Volin, S~Charles
  Doret, Harley Hayden, C-S Pai, David~W Landgren, Douglas Denison, Tyler
  Killian, Richart~E Slusher, and Alexa~W Harter.
\newblock Reliable transport through a microfabricated {X}-junction
  surface-electrode ion trap.
\newblock {\em New Journal of Physics}, 15(3):033004, 2013.

\bibitem{Ospelkaus2011}
C.~Ospelkaus, U.~Warring, Y.~Colombe, K.~R. Brown, J.~M. Amini, D.~Leibfried,
  and D.~J. Wineland.
\newblock Microwave quantum logic gates for trapped ions.
\newblock {\em Nature}, 476(7359):181--184, Aug 2011.

\bibitem{Shappert13}
C~M Shappert, J~T Merrill, K~R Brown, J~M Amini, C~Volin, S~C Doret, H~Hayden,
  C-S Pai, K~R Brown, and A~W Harter.
\newblock Spatially uniform single-qubit gate operations with near-field
  microwaves and composite pulse compensation.
\newblock {\em New Journal of Physics}, 15(8):083053, 2013.

\bibitem{Merrill11}
J~True Merrill, Curtis Volin, David Landgren, Jason~M Amini, Kenneth Wright,
  S~Charles Doret, C-S Pai, Harley Hayden, Tyler Killian, Daniel Faircloth,
  Kenneth~R Brown, Alexa~W Harter, and Richart~E Slusher.
\newblock Demonstration of integrated microscale optics in surface-electrode
  ion traps.
\newblock {\em New Journal of Physics}, 13(10):103005, 2011.

\bibitem{Wesenberg08}
J.~H. Wesenberg.
\newblock Electrostatics of surface-electrode ion traps.
\newblock {\em Phys. Rev. A}, 78:063410, Dec 2008.

\bibitem{Mehta2016}
Karan~K. Mehta, Colin~D. Bruzewicz, Robert McConnell, Rajeev~J. Ram, Jeremy~M.
  Sage, and John Chiaverini.
\newblock Integrated optical addressing of an ion qubit.
\newblock {\em Nat Nano}, 11(12):1066--1070, Dec 2016.
\newblock Letter.

\bibitem{Kunert14}
P.J. Kunert, D.~Georgen, L.~Bogunia, M.T. Baig, M.A. Baggash, M.~Johanning, and
  Ch. Wunderlich.
\newblock A planar ion trap chip with integrated structures for an adjustable
  magnetic field gradient.
\newblock {\em Applied Physics B}, 114(1-2):27--36, 2014.

\bibitem{Arrington13}
Christian~L. Arrington, Kyle~S. McKay, Ehren~D. Baca, Jonathan~J. Coleman, Yves
  Colombe, Patrick Finnegan, Dustin~A. Hite, Andrew~E. Hollowell, Robert
  J\"ordens, John~D. Jost, Dietrich Leibfried, Adam~M. Rowen, Ulrich Warring,
  Martin Weides, Andrew~C. Wilson, David~J. Wineland, and David~P. Pappas.
\newblock Micro-fabricated stylus ion trap.
\newblock {\em Review of Scientific Instruments}, 84(8):085001, 2013.

\bibitem{Bloch05}
Immanuel Bloch.
\newblock Ultracold quantum gases in optical lattices.
\newblock {\em Nat Phys}, 1(1):23--30, Oct 2005.

\bibitem{Guise15}
Nicholas~D. Guise, Spencer~D. Fallek, Kelly~E. Stevens, K.~R. Brown, Curtis
  Volin, Alexa~W. Harter, Jason~M. Amini, Robert~E. Higashi, Son~Thai Lu,
  Helen~M. Chanhvongsak, Thi~A. Nguyen, Matthew~S. Marcus, Thomas~R. Ohnstein,
  and Daniel~W. Youngner.
\newblock Ball-grid array architecture for microfabricated ion traps.
\newblock {\em Journal of Applied Physics}, 117(17):174901, 2015.

\bibitem{Guise14}
Nicholas~D. Guise, Spencer~D. Fallek, Harley Hayden, C-S Pai, Curtis Volin,
  K.~R. Brown, J.~True Merrill, Alexa~W. Harter, Jason~M. Amini, Lisa~M. Lust,
  Kelly Muldoon, Doug Carlson, and Jerry Budach.
\newblock In-vacuum active electronics for microfabricated ion traps.
\newblock {\em Review of Scientific Instruments}, 85(6), 2014.

\bibitem{Brown}
Ken Wright.
\newblock private communication.

\bibitem{Maunz16}
Peter Maunz.
\newblock {High Optical Access Trap 2.0}.
\newblock Technical report, Sandia National Laboratories, 01 2016.

\bibitem{Maunz16-2}
Peter Maunz.
\newblock High-fidelity two-qubit quantum gates in a scalable surface ion trap.
\newblock Southwest Quantum Information and Technology, 2016.

\bibitem{Martinez2016}
Esteban~A. Martinez, Christine~A. Muschik, Philipp Schindler, Daniel Nigg,
  Alexander Erhard, Markus Heyl, Philipp Hauke, Marcello Dalmonte, Thomas Monz,
  Peter Zoller, and Rainer Blatt.
\newblock Real-time dynamics of lattice gauge theories with a few-qubit quantum
  computer.
\newblock {\em Nature}, 534(7608):516--519, Jun 2016.
\newblock Letter.

\bibitem{Brandl16}
M.~F. Brandl, M.~W. van Mourik, L.~Postler, A.~Nolf, K.~Lakhmanskiy, R.~R.
  Paiva, S.~Moller, N.~Daniilidis, H.~Haffner, V.~Kaushal, T.~Ruster,
  C.~Warschburger, H.~Kaufmann, U.~G. Poschinger, F.~Schmidt-Kaler,
  P.~Schindler, T.~Monz, and R.~Blatt.
\newblock Cryogenic setup for trapped ion quantum computing.
\newblock {\em Review of Scientific Instruments}, 87(11):113103, 2016.

\bibitem{Ghadimi16}
Moji Ghadimi, Valdis Blums, Benjamin~G. Norton, Paul~M. Fisher, Steven~C.
  Connell, Jason~M. Amini, Curtis Volin, Harley Hayden, Chien-Shing Pai, David
  Kielpinski, Mirko Lobino, and Erik~W. Streed.
\newblock Scalable ion-photon quantum interface based on integrated diffractive
  mirrors.
\newblock {\em npj Quantum Information}, 3:1, 2016.

\bibitem{Monroe1164}
C.~Monroe and J.~Kim.
\newblock Scaling the ion trap quantum processor.
\newblock {\em Science}, 339(6124):1164--1169, 2013.

\bibitem{Matsukevich08}
D.~N. Matsukevich, P.~Maunz, D.~L. Moehring, S.~Olmschenk, and C.~Monroe.
\newblock Bell inequality violation with two remote atomic qubits.
\newblock {\em Phys. Rev. Lett.}, 100:150404, Apr 2008.

\bibitem{Mount2015}
Emily Mount, Daniel Gaultney, Geert Vrijsen, Michael Adams, So-Young Baek, Kai
  Hudek, Louis Isabella, Stephen Crain, Andre van Rynbach, Peter Maunz, and
  Jungsang Kim.
\newblock Scalable digital hardware for a trapped ion quantum computer.
\newblock {\em Quantum Information Processing}, pages 1--18, 2015.

\bibitem{nist_fpga}
National~Institute Standards and Technology (NIST).
\newblock Real-time infrastructure for quantum physics {(ARTIQ)}.
\newblock
  https://www.nist.gov/news-events/news/2015/01/open-source-software-quantum-information,
  2016.

\bibitem{sandia_fpga}
P.~Maunz.
\newblock Sandia {National} {Laboratory}.
\newblock https://github.com/pyIonControl, 2016.

\bibitem{Lin2015}
Guin-Dar Lin and L.-M. Duan.
\newblock Sympathetic cooling in a large ion crystal.
\newblock {\em Quantum Information Processing}, pages 1--15, 2015.

\bibitem{Kielpinski2015}
D.~Kielpinski, C.~Volin, E.~W. Streed, F.~Lenzini, and M.~Lobino.
\newblock Integrated optics architecture for trapped-ion quantum information
  processing.
\newblock {\em Quantum Information Processing}, pages 1--24, 2015.

\bibitem{Eltony2016}
Amira~M. Eltony, Dorian Gangloff, Molu Shi, Alexei Bylinskii, Vladan
  Vuleti{\'{c}}, and Isaac~L. Chuang.
\newblock Technologies for trapped-ion quantum information systems.
\newblock {\em Quantum Information Processing}, pages 1--33, 2016.

\end{thebibliography}
\bibliographystyle{unsrt}

\end{document}